\documentclass[11pt, aps, nofootinbib, prd]{revtex4-1}

\usepackage{geometry}
\usepackage{latexsym}
\usepackage{graphicx}
\usepackage{amsmath}
\usepackage{amssymb}
\usepackage{color}
\usepackage{mathtools}
\usepackage[colorlinks=true,backref=false,linktocpage=true,citecolor=blue,urlcolor=blue,linkcolor=blue,pdfpagemode=UseOutlines]{hyperref}
\usepackage{braket}
\usepackage{adjustbox}

\newcommand{\be}{\begin{equation}}
\newcommand{\ee}{\end{equation}}
\newcommand{\bea}{\begin{eqnarray}}
\newcommand{\eea}{\end{eqnarray}}

\newcommand{\RomatreINFN}{Istituto Nazionale di Fisica Nucleare, Sezione di Roma Tre,\\ Via della Vasca Navale 84, I-00146 Rome, Italy}
\newcommand{\LaSapienza}{Physics Department, University of Roma ``La Sapienza'' and INFN, Sezione di Roma,\\ Piazzale Aldo Moro 5, 00185 Roma, Italy}
\newcommand{\Pisa}{Dipartimento di Fisica dell'Universit\`a di Pisa and INFN, Sezione di Pisa,\\ Largo Bruno Pontecorvo 3, I-56127 Pisa, Italy}
\newcommand{\SNS}{Scuola Normale Superiore, Piazza dei Cavalieri 7, 56126 Pisa, Italy}
\newcommand{\INFNPisa}{INFN, Sezione di Pisa, Largo Bruno Pontecorvo 3, I-56127 Pisa, Italy}

\begin{document}

\title{$\vert V_{cb} \vert$, Lepton Flavour Universality and $SU(3)_F$ symmetry breaking\\ in $B_s \to D_s^{(*)} \ell \nu_\ell$ decays through unitarity and lattice QCD}

\author{G.\,Martinelli}\affiliation{\LaSapienza}
\author{M.\,Naviglio}\affiliation{\Pisa}
\author{S.\,Simula}\affiliation{\RomatreINFN}
\author{L.\,Vittorio}\affiliation{\SNS}\affiliation{\INFNPisa}

\begin{abstract}
In addition to the well-known $B \to D^{(*)} \ell \nu_\ell$ decays, semileptonic $B_s \to D_s^{(*)} \ell \nu_\ell$ processes offer the possibility to determine the Cabibbo-Kobayashi-Maskawa (CKM) matrix element $\vert V_{cb}\vert$. We implement the Dispersive Matrix (DM) approach to describe the hadronic Form Factors (FFs) for the $B_s \to D_s^{(*)}$ transition in the whole kinematical range, starting from recent Lattice QCD computations at large values of the 4-momentum transfer. We extract $\vert V_{cb} \vert$ from the experimental data, obtaining $\vert V_{cb} \vert \cdot 10^3 = (41.7 \pm 1.9)$ from $B_s \to D_s \ell \nu_\ell$ and $\vert V_{cb} \vert \cdot 10^3 =(40.7 \pm 2.4)$ from $B_s \to D_s^* \ell \nu_\ell$ decays. After averaging with the values of $\vert V_{cb} \vert$ obtained from the $B \to D^{(*)}$ channels\,\cite{Martinelli:2021onb,Martinelli:2021myh} we get $\vert V_{cb} \vert \cdot 10^3 =(41.2 \pm 0.8)$, which is compatible with the most recent inclusive estimate $\vert V_{cb} \vert_{\rm{incl}} \cdot 10^3 = 42.16 \pm 0.50$\,\cite{Bordone:2021oof} at the $1 \sigma$ level. In addition we test the Lepton Flavour Universality (LFU) by computing the $\tau / \ell$ ratios of the total decay rates (where $\ell$ is a light lepton), obtaining $R(D_s) = 0.298\,(5)$ and $R(D_s^*)= 0.250\,(6)$. We also address the issue of  the $SU(3)_F$ symmetry breaking by comparing the hadronic FFs entering the semileptonic $B \to D^{(*)}$ and $B_s \to D_s^{(*)}$ channels.
\end{abstract}

\maketitle

\newpage

\section{Introduction}
\label{sec:introduction}

Along many years the tension between the value of the  Cabibbo-Kobayashi-Maskawa (CKM) matrix element $\vert V_{cb} \vert$ determined from exclusive or inclusive $B$-meson decays remained as a puzzle. 
Recently\,\cite{Martinelli:2021onb, Martinelli:2021myh} new estimates of $\vert V_{cb} \vert$ have been obtained from exclusive semileptonic $B \to D^{(*)} \ell \nu_\ell$ decays by using a novel unitarity description of the relevant hadronic Form Factors (FFs)\,\cite{Lellouch:1995yv, DiCarlo:2021dzg} and the non-perturbative determination of the dispersive bounds corresponding to the $b \to c$ transition\,\cite{Martinelli:2021frl}. After an accurate estimate of the uncertainties we found
\bea
     \label{eq:Vcb_BD}
     \vert V_{cb} \vert \cdot 10^3 & = & (41.0 \pm 1.2) \qquad \mbox{from~} B \to D \ell \nu_\ell ~ \mbox{decays\,} \text{\cite{Martinelli:2021onb}} ~ \\[2mm]
     \label{eq:Vcb_BDstar}
                                                  & = & (41.3 \pm 1.7) \qquad \mbox{from~} B \to D^* \ell \nu_\ell ~ \mbox{decays\,} \text{\cite{Martinelli:2021myh}} ~ , ~
\eea
which are compatible with the most recent inclusive result $\vert V_{cb} \vert_{\rm{incl}} \cdot 10^3 = 42.16 \pm 0.50$\,\cite{Bordone:2021oof} at the $1 \sigma$ level. 

There are, however, other transitions that allow us to obtain the exclusive value of $\vert V_{cb} \vert$, like the semileptonic $B_s \to D_s^{(*)} \ell \nu_\ell$ decays. 
These transitions are very interesting since both LQCD\,\cite{McLean:2019qcx, Harrison:2021tol} and experimental\,\cite{Aaij:2020hsi,LHCb:2020hpv,LHCb:2021qbv} data became recently available.  Our aim is to examine the $B_s \to D_s^{(*)}$ transitions through the Dispersive Matrix (DM) method using the non-perturbative determination of the dispersive bounds for the $b \to c$ transition carried our in Ref.\,\cite{Martinelli:2021frl}. 
As done for the analysis of the exclusive $B \to D^{(*)}$ decays, we stress that we only use LQCD computations of the FFs to determine their shape as a function of the 4-momentum transfer, while the experimental data are used to obtain the final exclusive determination of $\vert V_{cb} \vert$. This fact also allows us to perform a pure, unbiased theoretical calculation of other quantities of phenomenological interest, like the $\tau / \ell$ ratios of total decay rates (where $\ell$ is an electron or a muon), crucial for testing the issue of Lepton Flavour Universality (LFU).

The possibility of another exclusive estimate of $\vert V_{cb} \vert$ is offered by the semileptonic $B_c \to J/\psi \ell \nu_\ell$ decays. However, the study of this process is more challenging since also the spectator quark is a heavy, charm quark. This implies that we need to compute the 2-loop contributions to the relevant susceptibilities (see Ref.\,\cite{Martinelli:2021frl}). This will be the subject of a future specific work. 

Using the DM method for describing the hadronic FFs we perform three different analyses to extract $\vert V_{cb} \vert$ from the experimental data\,\cite{Aaij:2020hsi,LHCb:2020hpv,LHCb:2021qbv}, obtaining 
\bea
     \label{eq:Vcb_BsDs_intro}
     \vert V_{cb} \vert \cdot 10^3 & = & (41.7 \pm 1.9) \qquad \mbox{from~} B_s \to D_s \ell \nu_\ell ~ \mbox{decays} ~ \\[2mm]
     \label{eq:Vcb_BsDsstar_intro}
                                                  & = & (40.7 \pm 2.4) \qquad \mbox{from~} B_s \to D_s^* \ell \nu_\ell ~ \mbox{decays} ~ . ~
\eea
After averaging with the values\,(\ref{eq:Vcb_BD})-(\ref{eq:Vcb_BDstar}) obtained from the $B \to D^{(*)}$ channels we get the result
\be
     \label{eq:Vcb_all}
     \vert V_{cb} \vert \cdot 10^3 =(41.2 \pm 0.8)  \qquad \mbox{from~} B_{(s)} \to D_{(s)}^{(*)} \ell \nu_\ell ~ \mbox{decays} ~ , ~ 
\ee
which is compatible with the inclusive estimate $\vert V_{cb} \vert_{\rm{incl}} \cdot 10^3 = 42.16 \pm 0.50$\,\cite{Bordone:2021oof} at the $1 \sigma$ level.

The paper is organized as follows. In Section \,\ref{sec:FFs} we describe the state of the art of the computations of the FFs on the lattice and apply the DM method to the FFs entering the semileptonic $B_s \to D_s^{(*)} \ell \nu$ decays. In Section\,\ref{sec:applications} we extract $\vert V_{cb} \vert$ from the experimental data for both $B_s \to D_s$ and $B_s \to D_s^*$ channels. Then, we compute pure, unbiased theoretical values of the LFU ratios of total decay rates, $R(D_s^{(*)})$, as well as of the polarization observables $P_{\tau}^{s}$ and $F_L^{s}$. In Section\,\ref{sec:SU3_F} we compare the hadronic FFs entering $B \to D^{(*)}$ and  $B_s \to D_s^{(*)}$ decays and address the issue of $SU(3)_F$ symmetry breaking. Section\,\ref{sec:conclusions} summarizes our conclusions.

\section{The Form Factors entering the semileptonic $B_s \to D_s^{(*)} \ell \nu_\ell$ decays}
\label{sec:FFs}

In this Section we present the theoretical expressions of the differential decay widths for the semileptonic $B_s \to D_s^{(*)} \ell \nu_\ell$ decays in terms of the relevant hadronic FFs within the SM. Then, we present the state of the art of the LQCD computations of the FFs and, through the DM method, we study their shape in the whole kinematical range using as inputs only the lattice results at large values of the 4-momentum transfer.

\subsection{The differential decay widths in the SM}
\label{sec:widths}

For the decay in a pseudoscalar meson (i.e.~for the $B_s \to D_s$ transition), the relevant hadronic matrix element reads
\be
    \label{eq:FINALmatrelem}
     \Braket{D_s(p_{D_s})| V^{\mu} |B_s(p_{B_s})} = f_{+}^s(q^2)\left(p_{B_s}^{\mu}+p_{D_s}^{\mu} - \frac{m_{B_s}^2-m_{D_s}^2}{q^2}q^{\mu}\right) +
                                                                                  f_{0}^s(q^2) \frac{m_{B_s}^2-m_{D_s}^2}{q^2} q^{\mu},
\ee
where $V^\mu \equiv \bar{b} \gamma^\mu c$ is the hadronic weak vector current for the $b \to c$ transition and $q = p_{B_s} - p_{D_s}$ is the 4-momentum transfer.
The vector $f_+^s(q^2)$ and the scalar $f_0^s(q^2)$ FFs are related at $q^2 = 0$ by a kinematical constraint (KC). 
In terms of the recoil variable $w \equiv v_{B_s} \cdot v_{D_s}$ with $v_{B_s(D_s)}$ being the meson 4-velocities, the KC reads as
\be
    \label{eq:KC0}
    f_+^s(w_{max}) = f_0^s(w_{max}) ~ , ~
\ee 
where $w_{max} \equiv  (m_{B_s}^2 + m_{D_s}^2)  / (2 m_{B_s} m_{D_s})$.
The differential decay width $d\Gamma / dq^2$ is given by
\bea
    \label{eq:GammaBsDs}
    && \frac{d\Gamma}{dq^2}(B_s \to D_s\ell \nu_\ell)  = \frac{G_F^2 \vert V_{cb} \vert^2 \eta_{EW}^2}{24\pi^3} \left(1-\frac{m_{\ell}^2}{q^2}\right)^2 \\[2mm]
    && \hspace{2cm} \cdot \left[\vert \vec{p}_{D_s}\vert^3 \left(1+\frac{m_{\ell}^2}{2q^2}\right) \vert f_+^s(q^2) \vert^2 + 
         m_{B_s}^2 \vert \vec{p}_{D_s} \vert \left( 1-\frac{m_{D_s}^2}{m_{B_s}^2}\right)^2 \frac{3m_{\ell}^2}{8q^2} \vert f_0^s(q^2) \vert^2\right] ~ , ~ \nonumber
\eea
where $G_F$ is the Fermi constant, $\vec{p}_{D_s}$ the 3-momentum of the $D_s$-meson in the rest frame of the $B_s$-meson, $\eta_{EW} \simeq 1.0066$ is the leading electroweak correction\,\cite{Sirlin:1981ie} and $m_\ell$ is the mass of the final charged lepton. 

For the decay into a vector meson (i.e.~for the $B_s \to D_s^*$ transition) both the vector $V^{\mu}$ and the axial-vector $A^{\mu}\equiv \bar{b}\gamma^{\mu}\gamma^5c$ currents contribute to the amplitude of the process, namely
\bea
\label{eq:matrix_el_Dstar}
& & \langle D_s^{*} (p_{D_s^*}, \epsilon)| \bar{c} \gamma^\mu \left(1 \mp \gamma_5 \right) b |\bar{B_s}(p_{B_s}) \rangle =
       - \frac{1}{m_{B_s} + m_{D_s^*}}  \varepsilon^{\mu}_{\alpha \beta \gamma} \epsilon^{*\alpha} P^\beta q^\gamma V^s(q^2) ~ \nonumber \\[2mm]
& &\hspace{2.5cm} \pm \, i \, \frac{2 m_{D_s^*}}{q^2} (\epsilon^* \cdot q ) \, q^\mu A_0^s(q^2) 
                            \mp \, i \, \left[ \epsilon^{*\mu} - \frac{\epsilon^* \cdot q}{m_{B_s} - m_{D_s^*}} P^\mu \right] A_1^s (q^2) ~ \\[2mm]
& & \hspace{2.5cm} \mp \, i \, \frac{ 2 m_{D_s^*}}{q^2} (\epsilon^* \cdot q) \left[\frac{q^2}{m_{B_s}^2 - m_{D_s^*}^2} P^\mu - q^\mu \right] A_3^s(q^2) ~ , ~ 
       \nonumber
\eea
where $2 m_{D_s^*}A_3^s(q^2) \equiv \left[ (m_{B_s} + m_{D_s^*}) A_1^s(q^2)-  (m_{B_s} - m_{D_s^*}) A_2^s(q^2) \right]$, $P \equiv p_{D_s^*} + p_{B_s}$ and $\epsilon$ is the polarisation vector of the final $D_s^*$-meson. 
The FFs $V^s$, $A_1^s$, $A_2^s$ and $A_0^s$ are related to those corresponding to a definite spin-parity\,\cite{Boyd:1995sq, Boyd:1995cf, Boyd:1997kz} by
\bea
    \label{eq:ff_BtoDstar}
    V^s(w) & = & \frac{m_{B_s}+m_{D_s^*}}{2} g^s(w)  ~ \\[2mm]
    \label{eq:ff_BtoDstar2}
    A_1^s(w) & = &  \frac{f^s(w)}{m_{B_s} + m_{D_s^{*}}}  ~ \\ [2mm]
    \label{eq:ff_BtoDstar3}
    A_2^s(w) & = & \frac{1}{2} \frac{m_{B_s} + m_{D_s^*}}{m_{B_s} m_{D_s^*}}  \frac{1}{w^2 - 1}
                              \left[ \left(w - \frac{m_{D_s^*}}{m_{B_s}}\right) f^s(w) - \frac{\mathcal{F}_1^s(w)}{m_{B_s}} \right]  ~ \\[2mm]  
   \label{eq:ff_BtoDstar4}
   A_0^s(w) & = & \frac{1}{2} \frac{m_{B_s} + m_{D_s^*}}{\sqrt{m_{B_s} m_{D_s^*}}} P_1^s(w)  ~ , ~ 
\eea
where $w \equiv v_{B_s} \cdot v_{D_s^*}$ is the recoil variable with $v_{B_s(D_s^*)}$ being the meson 4-velocities. 
The FFs should satisfy two KCs: the first one applies at minimum recoil
\be
    \label{eq:KC1}
    \mathcal{F}^s_1(w=1) = (m_{B_s} - m_{D_s^*})f^s(w=1) ~ , ~
\ee
while the second one holds at maximum recoil
\be
    \label{eq:KC2}
    P_1^s (w_{max}^*) = \frac{\mathcal{F}_1^s(w_{max}^*)}{(1 + w_{max}^*)(m_{B_s} - m_{D_s^*}) \sqrt{m_{B_s} m_{D_s^*}}} ~ , ~
\ee
where $w_{max}^* \equiv  (m_{B_s}^2 + m_{D_s^*}^2)  / (2 m_{B_s} m_{D_s^*})$. Note that another notation for the pseudoscalar FF, $\mathcal{F}^{s}_2(w)$, can be found in the literature\,\cite{Boyd:1995sq, Boyd:1995cf, Boyd:1997kz}. The two notations differ from each other only by a simple kinematical factor, namely $P_1^{s}(w) =  \mathcal{F}^{s}_2(w) \sqrt{m_{B_s} m_{D_s^*}} / (m_{B_s} +  m_{D_s^*})$.

Finally, from the matrix element\,(\ref{eq:matrix_el_Dstar}) it follows that in the limit of massless leptons the differential decay width $d\Gamma / dq^2$ is given by 
\bea
    \label{eq:GammaBsDsstar}
    && \frac{d\Gamma}{dq^2}(B_s \to D_s^* \ell \nu_\ell) = \frac{G_F^2 \vert V_{cb} \vert^2  \eta_{EW}^2 m_{D_s^*}}{96 \pi ^3 m_{B_s}^2} \sqrt{w^2-1}  ~ \\[2mm]
    && \hspace{2cm} \cdot \left\{ \left[ \mathcal{F}^s_1(w) \right]^2  + 2 \, q^2 \left( \left[ f^s(w) \right]^2 + 
          m_{B_s}^2 m_{D_s^*}^2 \left( w^2 - 1 \right) \left[ g^s(w) \right]^2 \right) \right\} ~ , ~ \nonumber
\eea
where
\be
   \label{eq:q2w}
   q^2 = m_{B_s}^2 + m_{D_s^*}^2 - 2 m_{B_s} m_{D_s^*} w = 2 m_{B_s} m_{D_s^*} \left( w_{max}^* - w \right) ~ . ~ \nonumber
\ee

\subsection{State of the art of the LQCD computations of the FFs}
\label{sec:LQCD}

The FFs entering semileptonic $B_s \to D_s^{(*)} \ell \nu_\ell$ decays have been computed on the lattice by the HPQCD Collaboration in Refs.\,\cite{McLean:2019qcx, Harrison:2021tol}. In Ref.\,\cite{McLean:2019qcx}, the authors have made available the results of Bourrely-Caprini-Lellouch (BCL)\,\cite{Bourrely:2008za} fits of the FFs extrapolated to the physical $b$-quark point and to the continuum limit. In Ref.\,\cite{Harrison:2021tol}, instead, the authors have used a different $z$-expansion parameterisation of the FFs. The mean values and the covariance matrix of the $z$-expansion coefficients in the continuum limit can be found in Ref.\,\cite{Harrison:2021tol}. 
The fits of Refs.\,\cite{McLean:2019qcx, Harrison:2021tol} provide the FFs in the whole kinematical range, but we want to use our DM method in order to extrapolate the shape of the FFs in the whole kinematical range, minimizing the impact of any assumption about the momentum dependence of the FFs. Therefore, we select only three values of $q^2$ in the high-$q^2$ regime, namely at $q^2 \approx \{ 0.7, 0.85, 1.0 \} \cdot q_{max}^2$, where $q_{max}^2 = 11.6$ GeV$^2$ for the $B_s \to D_s \ell \nu_\ell$ decays and $q_{max}^2 = 10.6$ GeV$^2$ for the $B_s \to D_s^* \ell \nu_\ell$ decays. We have explicitly checked that our results for the FFs in the whole kinematical range (see Section\,\ref{sec:applications_DM}) do not depend upon the specific choice of the locations of the input FF data for both the $B_s \to D_s \ell \nu_\ell$ and the $B_s \to D_s^* \ell \nu_\ell$ decays.
 
Then, from the marginalized values of the BCL or $z$-expansion coefficients we reconstruct the FFs in the high-$q^2$ regime in order to use them as inputs for the DM method. 
The mean values and uncertainties of the LQCD inputs are collected in Table\,\ref{tab:LQCDBsDs} for the $B_s \to D_s \ell \nu_\ell$ decays and in Table\,\ref{tab:LQCDBsDsstar} for the $B_s \to D_s^* \ell \nu_\ell$ decays. 

\begin{table}[htb!]
\renewcommand{\arraystretch}{1.1}
\begin{center}
\begin{tabular}{||c||c||c||}
\hline
~ $q^2$ (GeV$^2$) ~ & ~ $f_+^s(B_s \to D_s)$ ~ & ~ $f_0^s(B_s \to D_s)$ ~\\
\hline
~$8.5$  & ~ 1.021(28) ~ & ~ 0.834(12) ~\\
$10.0$  & ~ 1.108(34) ~ & ~ 0.873(13) ~\\
$11.6$  & ~ 1.209(41) ~ & ~ 0.917(15) ~ \\
\hline
\end{tabular}
\caption{\it \small Values of the vector $f_+^s(q^2)$ and scalar  $f_0^s(q^2)$ FFs for the $B_s \to D_s \ell \nu_\ell$ decays, evaluated at $q^2 = \{ 8.5,10.0,11.6 \} $ GeV$^2$ using the BCL fit computed by the HPQCD Collaboration in Ref.\,\cite{McLean:2019qcx}.}
\label{tab:LQCDBsDs}
\end{center}
\renewcommand{\arraystretch}{1.0}
\end{table}

\begin{table}[htb!]
\renewcommand{\arraystretch}{1.1}
\begin{center}
\begin{tabular}{||c||c||c||c||c||}
\hline
~ $q^2~\mbox{(GeV}^2)$ ~ & ~ $f^s(B_s \to D_s^*)$ ~ & ~ $g^s(B_s \to D_s^*)$ ~ & ~ $\mathcal{F}_1^s(B_s \to D_s^*)$ ~ & ~ $P_1^s(B_s \to D_s^*)$ ~\\
\hline
~$7.1$  & ~ 5.40(22) ~ & ~ 0.341(35) ~ & ~ 18.01(75) ~ & ~ 0.781(42) ~\\
~$8.9$  & ~ 5.73(22) ~ & ~ 0.369(38) ~ & ~ 18.91(70) ~ & ~ 0.861(45) ~\\
$10.6$  & ~ 6.09(22) ~ & ~ 0.401(44) ~  & ~ 19.81(73) ~ & ~ 0.949(51) ~\\
\hline
\end{tabular}
\caption{\it \small Values of the FFs $f^s, g^s, \mathcal{F}_1^s$ and $P_1^s$ for the $B_s \to D_s^* \ell \nu_\ell$ decays, evaluated at $q^2 = \{ 7.1, 8.9, 10.6 \}$ GeV$^2$ using the $z$-expansion fit computed by the HPQCD Collaboration in Ref.\,\cite{Harrison:2021tol}. The FF $P_1^s$ is dimensionless, while the FFs $f^s, g^s$ and $\mathcal{F}_1^s$ are given in units of GeV, GeV$^{-1}$ and GeV$^2$, respectively.}
\label{tab:LQCDBsDsstar}
\end{center}
\renewcommand{\arraystretch}{1.0}
\end{table}

\subsection{The DM method}
\label{sec:DM}

We now briefly recall the main features of the DM method applied to the description of a generic FF $f(q^2)$ with definite spin-parity.

Let us consider a set of $N$ values of the FF, $\{ f \} = \{ f(z_j) \}$ with $j = 1, 2, ..., N$, where $z$ is the conformal variable
\be
    \label{eq:z}
    z(q^2) \equiv \frac{\sqrt{t_+ - q^2} - \sqrt{t_+ - t_-}}{\sqrt{t_+ - q^2} + \sqrt{t_+ - t_-}} ~ 
\ee
with $t_\pm \equiv (m_{B_s} \pm m_{D_s^{(*)}})^2$ in the cases of our interest and $z_j \equiv z(q_j^2)$.
Then, the FF at a generic value of $z = z(q^2)$ is bounded by unitarity, analyticity and crossing symmetry to be in the range\,\cite{DiCarlo:2021dzg}
\be
  \beta(z) - \sqrt{\gamma(z)} \leq f(z) \leq \beta(z) + \sqrt{\gamma(z)} ~ , ~
    \label{eq:bounds}
\ee 
where 
\bea
      \label{eq:beta_final}
      \beta(z) & \equiv & \frac{1}{\phi(z, q_0^2) d(z)} \sum_{j = 1}^N f(z_j) \phi(z_j, q_0^2) d_j \frac{1 - z_j^2}{z - z_j} ~ , ~ \\
      \label{eq:gamma_final}
      \gamma(z) & \equiv &  \frac{1}{1 - z^2} \frac{1}{\phi^2(z, q_0^2) d^2(z)} \left[ \chi(q_0^2) - \chi_{\{f\}}^{DM}(q_0^2) \right] ~ , ~ \\
      \label{eq:chiDM}
      \chi_{\{f\}}^{DM}(q_0^2) & \equiv & \sum_{i, j = 1}^N f(z_i)f(z_j) \phi(z_i, q_0^2)  \phi(z_j, q_0^2) d_i d_j \frac{(1 - z_i^2) (1 - z_j^2)}{1 - z_i z_j} ~ 
\eea
with
\be
    \label{eq:dcoef}
    d(z) \equiv \prod_{m = 1}^N \frac{1 - z z_m}{z - z_m} ~ , ~ \qquad d_j  \equiv \prod_{m \neq j = 1}^N \frac{1 - z_j z_m}{z_j - z_m} ~ . ~
\ee
In the above Equations $\chi(q_0^2)$ is the dispersive bound, evaluated at an auxiliary value $q_0^2$ of the squared 4-momentum transfer using suitable two-point correlators, and $\phi(z, q_0^2)$ is a kinematical function appropriate for the given form factor\,\cite{Boyd:1997kz}. The kinematical function $\phi$ may be modified to include the contribution of the resonances below the pair production threshold $t_+$.

Unitarity is satisfied only when $\gamma(z) \geq 0$, which implies
\be
    \label{eq:UTfilter}
    \chi(q_0^2) \geq \chi_{\{f\}}^{DM}(q_0^2) ~ . ~
\ee
Since $\chi_{\{f\}}^{DM}(q_0^2)$ does not depend on $z$, Eq.\,(\ref{eq:UTfilter}) is either never verified or always verified for any value of $z$.
This leads to the first important feature of our method: the DM {\it unitarity filter}\,(\ref{eq:UTfilter}) represents a parameterization-independent implementation of unitarity for the given set of input values $\{ f \}$ of the FF.

We point out another important feature of the DM approach.
When $z$ coincides with one of the data points, i.e.~$z \to z_j$, one has $\beta(z) \to f(z_j)$ and $\gamma(z) \to 0$.
In other words the DM method reproduces exactly the given set of data points.
This leads to the second important feature of our method: the DM band given in Eq.~(\ref{eq:bounds}) is equivalent to the results of all possible fits that satisfy unitarity and at the same time reproduce exactly the input data.

The above features may not be shared by truncated parameterisations based on the $z$-expansion, like the Boyd-Grinstein-Lebed (BGL)\,\cite{Boyd:1997kz} or the BCL\,\cite{Bourrely:2008za} fits. Indeed, there is no guarantee that truncated parameterizations reproduce exactly the set of input data and, consequently, the fulfillment of the unitarity constraint may fictitiously depend upon the order of the truncation.

\subsection{Application of the DM method to the description of the FFs}
\label{sec:applications_DM}

We now apply the DM approach to the description of the FFs entering the semileptonic $B_s \to D_s^{(*)} \ell \nu_\ell$ decays. 

The non-perturbative values of the dispersive bounds corresponding to the $b \to c$ transition for the relevant channels with definite spin-parity have been computed on the lattice at $q_0^2 = 0$ in Ref.\,\cite{Martinelli:2021frl}. 
After subtraction of the contribution of bound states they are
\bea
     \label{eq:bound0+}
     \chi_{0^+}(0) & = & (7.58 \pm 0.59) \cdot 10^{-3} ~ , ~ \nonumber \\
      \label{eq:bound1-}
     \chi_{1^-}(0) & = & (5.84 \pm 0.44) \cdot 10^{-4} ~ \mbox{GeV}^{-2} ~ , ~ \\
     \label{eq:bound0-}
     \chi_{0^-}(0) & = & (21.9 \pm 1.9) \cdot 10^{-3} ~ , ~\nonumber  \\     
     \label{eq:bound1+}
     \chi_{1^+}(0) & = & (4.69 \pm 0.30) \cdot 10^{-4} ~ \mbox{GeV}^{-2} ~ ,~ \nonumber
\eea
The kinematical functions associated to the semileptonic FFs reads\,\cite{Boyd:1997kz}
\bea
    \label{eq:phif0}
    \phi_{f_0^s}(z, 0) & = & 2 r (1 - r^2) \sqrt{\frac{2 n_I}{ \pi}} \frac{(1-z^2) \sqrt{1 - z}}{\left[ (1 + r)(1 - z) + 2 \sqrt{r}(1 + z) \right]^4} ~ , ~  \\
    \label{eq:phif+}
    \phi_{f_+^s}(z, 0) & = & \frac{16 r^2}{m_{B_s}} \sqrt{\frac{2 n_I}{3 \pi}} \frac{(1 + z)^2 \sqrt{1 - z}}{\left[ (1 + r)(1 - z) + 2 \sqrt{r}(1 + z) \right]^5} ~ \nonumber
\eea 
with $r \equiv m_{D_s} / m_{B_s}$, and
\bea
    \label{eq:phif}
    \phi_{f^s}(z, 0) & = & 4 \frac{r_*}{m_{B_s}^2} \sqrt{\frac{n_I}{3\pi}} \, \frac{(1 + z)(1 - z)^{3/2}}{\left[ (1 + r_*)(1 - z) + 2 \sqrt{r_*}(1 + z) \right]^4} ~ , ~ \nonumber \\ 
    \label{eq:phig}
    \phi_{g^s}(z) & = & 16 r_*^2 \sqrt{\frac{n_I}{3\pi}} \frac{(1 + z)^{2}}{\sqrt{1 - z}\left[ (1 + r_*)(1 - z) + 2 \sqrt{r_*}(1 + z) \right]^4} ~ , ~ \nonumber \\
    \label{eq:phiF1}
    \phi_{\mathcal{F}_1^s}(z, 0) & = & 2 \frac{r_*}{m_{B_s}^3} \sqrt{\frac{2n_I}{3\pi}} \frac{(1 + z)(1 - z)^{5/2}}{\left[ (1 + r_*)(1 - z) + 2 \sqrt{r_*}(1 + z) \right]^5} ~ , ~ \\
    \label{eq:phiP1}
    \phi_{P_1^s}(z, 0) & = &  8 (1 + r_*) r_*^{3/2} \sqrt{\frac{2n_I}{\pi}} \frac{(1 + z)^{2}}{\sqrt{1 - z}\left[ (1 + r_*)(1 - z) + 2 \sqrt{r_*}(1 + z) \right]^4} ~ \nonumber 
\eea
with $r_* \equiv m_{D_s^*} / m_{B_s}$.
In Eqs.\,(\ref{eq:phif0})-(\ref{eq:phiF1}) $n_I$ is a factor counting the number of spectator quarks and it is equal to $n_I = 1$ for the $B_s \to D_s^{(*)} \ell \nu_\ell$ decays\,\cite{Boyd:1997kz} .

The presence of resonances below the pair production threshold lead to the following modification of the kinematical function $\phi(z, 0)$\,\cite{Lellouch:1995yv}
\be
    \label{eq:poles}
    \phi(z, 0) \to \phi(z, 0) \cdot \prod_R \frac{z - z(m_R^2)}{1 - z \, z(m_R^2)} ~ , ~
\ee
where $m_R$ is the mass of the resonance $R$.
For the masses of the poles corresponding to $B_c^{(*)}$-mesons with different quantum numbers entering the various FFs we refer to Appendix A of Ref.\,\cite{McLean:2019qcx} for the $B_s \to D_s \ell \nu_\ell$ decays and to Table XII of Ref.\,\cite{Harrison:2021tol} for the $B_s \to D_s^* \ell \nu_\ell$ decays.

For the unitarity constraints (see Eq.\,(\ref{eq:UTfilter})) we consider
\begin{itemize}

\item in the case of the  $B_s \to D_s \ell \nu_\ell$ decays
\bea
   \label{eq:UTfilter1}
    \chi_{0^+}(0) & \geq & \chi_{\{ f_0^s \}}^{DM}(0) ~ , ~ \\
   \label{eq:UTfilter2}
    \chi_{1^-}(0) & \geq & \chi_{\{ f_+^s \}}^{DM}(0) ~ , ~ \nonumber
\eea

\item in the case of the  $B_s \to D_s^* \ell \nu_\ell$ decays
\bea
   \label{eq:UTfilter3}
    \chi_{1^-}(0) & \geq & \chi_{\{ g^s \}}^{DM}(0) ~ , ~ \nonumber \\
   \label{eq:UTfilter4}
    \chi_{1^+}(0) & \geq & \chi_{\{ f^s \}}^{DM}(0) + \chi_{\{ \mathcal{F}_1^s \}}^{DM}(0) ~ , ~ \\
   \label{eq:UTfilter5}
    \chi_{0^-}(0) & \geq & \chi_{\{ P_1^s \}}^{DM}(0) ~ \nonumber
\eea

\end{itemize}
and we examine the effect of the above unitarity filters on the bootstrap events generated through a multivariate Gaussian distribution based on the computations of the FFs by the HPQCD Collaboration for the  $B_s \to D_s^{(*)} \ell \nu_\ell$ decays (see Tables\,\ref{tab:LQCDBsDs}-\ref{tab:LQCDBsDsstar}).
For both decays 100$\%$ of the generated bootstraps survive to the unitarity filters and this holds as well for each separate spin-parity quantum channel.
The percentage of surviving events does not change even if we consider the combined filter $\chi_{1^-}(0) \geq \chi_{\{ f_+^s \}}^{DM}(0) + \chi_{\{ g^s \}}^{DM}(0)$, connecting the two decay processes.
The above results are not surprising. Indeed, the values of the dispersive bounds\,(\ref{eq:bound1-}) are very conservative ones, since they sum up the contribution of all spectator quarks, both the light $u$- and $d$-quarks as well as the strange and the charm quarks. 
Thus, assuming small $SU(3)_F$ breaking effects in the FFs, we consider the alternative case in which the kinematical functions\,(\ref{eq:phif0})-(\ref{eq:phiF1}) are evaluated using $n_I = 3$.
This is equivalent to leave unchanged all the kinematical functions, but to divide the dispersive bounds\,(\ref{eq:bound1-}) by three (see also Eqs.\,(\ref{eq:beta_final})-(\ref{eq:chiDM})).
The net result is that, as in the previous case, almost 100$\%$ of the generated bootstraps survive to the new unitarity filters.
The same holds as well for the application of the three KCs\,(\ref{eq:KC0}) and (\ref{eq:KC1})-(\ref{eq:KC2}).
Therefore, neither the skeptical nor the iterative procedures described in Refs.\,\cite{DiCarlo:2021dzg,Martinelli:2021onb,Martinelli:2021myh} need to be applied.

Notice that our choice of the conformal variable $z$ (see Eq.\,(\ref{eq:z})) may lead to the occurrence of branch points related to multiparticle production inside the unit circle $|z| =1$. However, following the approach of Ref.\,\cite{Boyd:1995sq}, we have verified that their impact on the dispersive bounds\,(\ref{eq:bound1-}), as well as on the input data of the semileptonic FFs, is expected to be small and well within the uncertainties.

\begin{figure}[htb!]
\begin{center}
\includegraphics[scale=0.50]{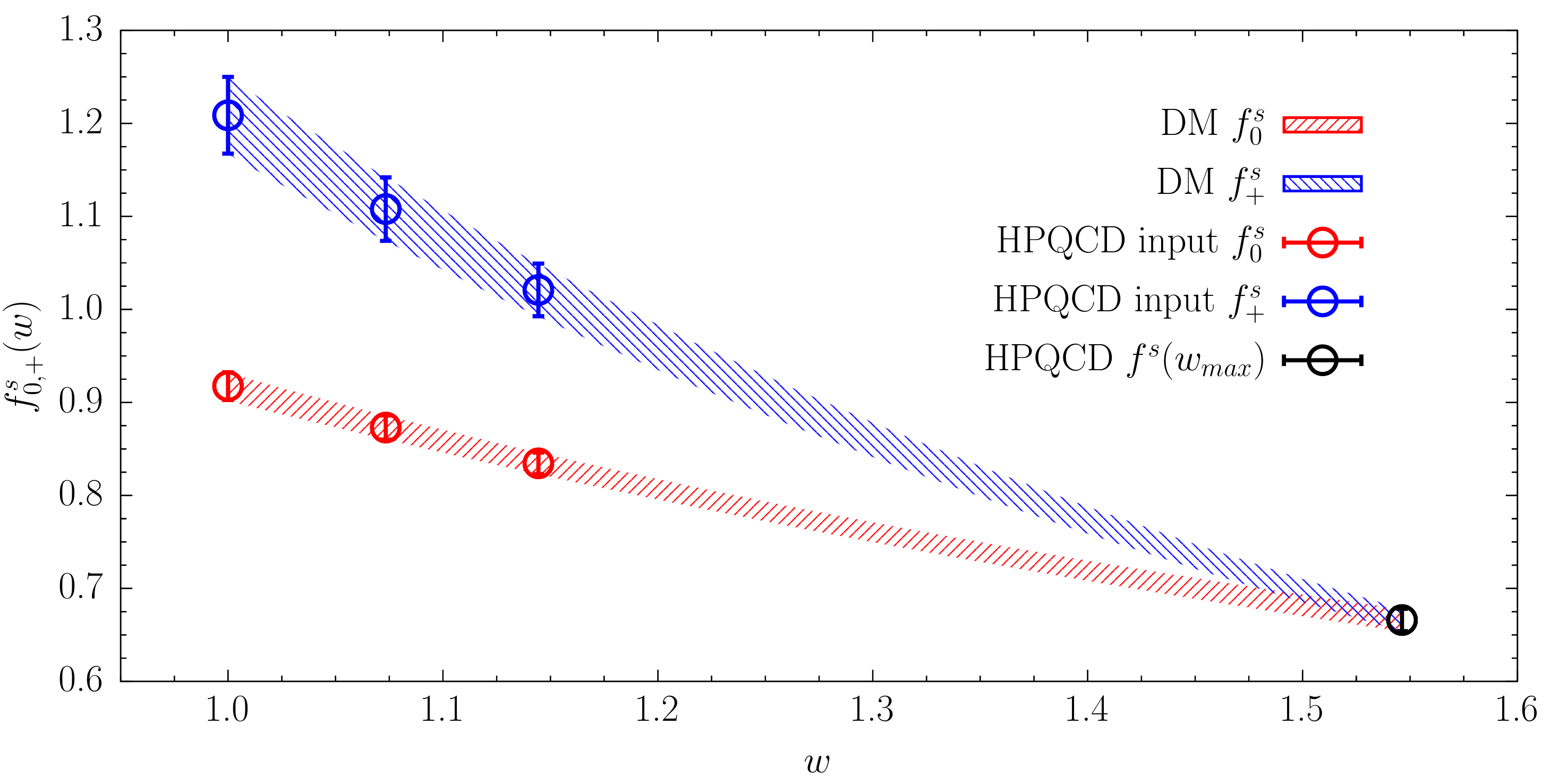}
\vspace{-0.5cm}
\caption{\it \small The bands of the scalar and vector FFs entering the $B_s \to D_s \ell \nu_\ell$ decays computed through the DM method versus the recoil variable $w$. The blue band represents $f_+^s(w)$, while the red one is $f_0^s(w)$. The blue and red circles are the values of the FFs obtained by the HPQCD Collaboration in Ref.\,\cite{McLean:2019qcx} and used as inputs for the DM method (see Table\,\ref{tab:LQCDBsDs}). The black circle represents the (common) value of the FFs at maximum recoil obtained by the HPQCD Collaboration in Ref.\,\cite{McLean:2019qcx}.}
\label{fig:FFMMBsDs}
\end{center}
\end{figure}

\begin{figure}[htb!]
\begin{center}
\includegraphics[scale=0.50]{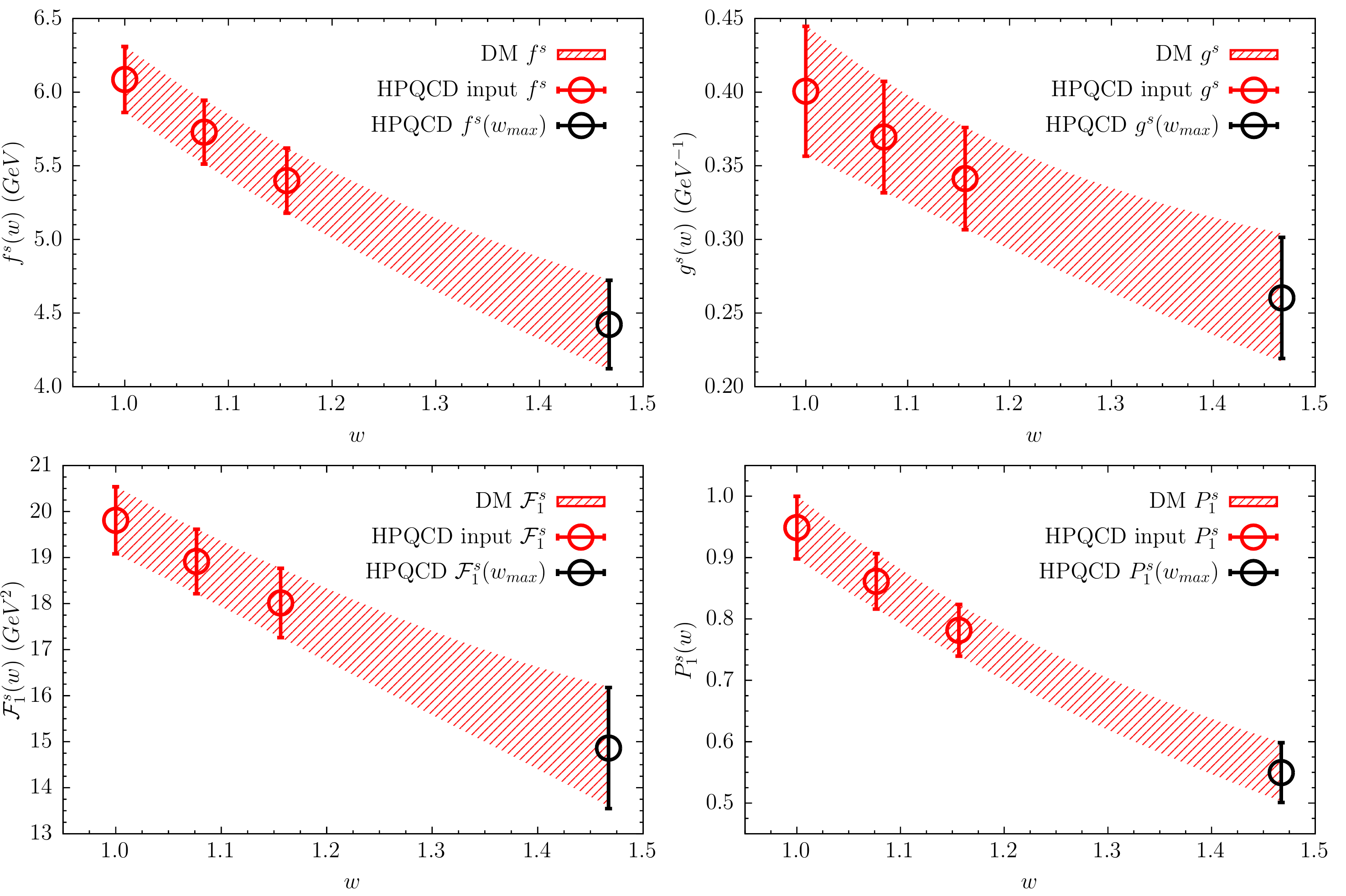}
\vspace{-0.5cm}
\caption{\it \small The bands of the four FFs entering the $B_s \to D_s^* \ell \nu$ decays, i.e.~$f^s(w)$, $g^s(w)$, $\mathcal{F}_1^s(w)$ and $P_1^s(w)$, computed through the DM method versus the recoil variable $w$. The red circles are the values of the FFs obtained by the HPQCD Collaboration in Ref.\,\cite{Harrison:2021tol} and used as inputs for the DM method (see Table\,\ref{tab:LQCDBsDsstar}). The black circles represent the values of the FFs at maximum recoil obtained by the HPQCD Collaboration in Ref.\,\cite{Harrison:2021tol}.}
\label{fig:FFMMBsDsstar}
\end{center}
\end{figure}

The DM bands of the FFs are shown in Figs.\,\ref{fig:FFMMBsDs} and \ref{fig:FFMMBsDsstar}.
The extrapolations of the FFs at maximum recoil, which are important for the phenomenological applications which will be discussed in Section~\ref{sec:applications}, read 
\be
    \label{eq:f+0_wmax}
    f_+^s(w_{max}) = f_0^s(w_{max}) = 0.666 \pm 0.012 ~ , ~ \\
 \ee
 and
 \bea
    \label{eq:f_wmax} 
     f^s(w_{max}^*) & = & 4.42 \pm 0.30 \,\rm{GeV},\\
    \label{eq:g_wmax}
    g^s(w_{max}^*) & = & 0.261 \pm 0.044 \,\rm{GeV}^{-1},\\
    \label{eqF1_wmax}
    \mathcal{F}^s_1(w_{max}^*) & = & 14.9 \pm 1.3 \,\rm{GeV}^2,\\
    \label{eq:P1_wmax}
     P^s_1(w_{max}^*) &= & 0.551 \pm 0.048 ~ , ~   
\eea
which are consistent with the values of the fits performed by the HPQCD Collaboration in Refs.\,\cite{McLean:2019qcx,Harrison:2021tol}, as shown in Figs.\,\ref{fig:FFMMBsDs}-\ref{fig:FFMMBsDsstar}. 

We remind that the DM bands of the FFs do not depend upon the specific choice of the locations of the input FF data for both the $B_s \to D_s \ell \nu_\ell$ and the $B_s \to D_s^* \ell \nu_\ell$ decays.

\section{Phenomenological applications to $B_s \to D_s^{(*)} \ell \nu_\ell$ decays}
\label{sec:applications}

In this Section we determine the value of $\vert V_{cb} \vert$ from $B_s \to D_s^{(*)} \ell \nu_\ell$ decays using the available experimental data\,\cite{Aaij:2020hsi,LHCb:2020hpv,LHCb:2021qbv} and the DM bands of the hadronic FFs determined in the previous Section. 
We apply the bin-per-bin strategy already adopted in Refs.\,\cite{Martinelli:2021onb, Martinelli:2021myh} to investigate the semileptonic $B \to D^{(*)} \ell \nu_\ell$ decays. We stress that in our approach the hadronic FFs (including their uncertainties) are determined exclusively by our fundamental theory of strong interactions, i.e.~QCD, while the experimental data are used only  to obtain the final exclusive determination of $\vert V_{cb} \vert$.
We also give pure theoretical estimates of the $\tau/\mu$ ratios of decay rates, which are crucial for testing LFU, and of various polarization observables. Finally, we investigate the issue of $SU(3)_F$ symmetry breaking in semileptonic $b \to c \ell \nu_\ell$ transitions.

\subsection{Determination of $\vert V_{cb} \vert$}
\label{sec:Vcb}

As far as the experimental measurements are concerned, the integrated branching fractions for both the $B_s \to D_s \ell \nu_\ell$ and the $B_s \to D_s^* \ell \nu_\ell$ processes have been determined in Ref.\,\cite{Aaij:2020hsi} and updated in Ref.\,\cite{LHCb:2021qbv}. This allows us to determine $\vert V_{cb} \vert$ for the two channels using the theoretical estimate of the branching ratios based on the DM bands of the semileptonic FFs, as illustrated in Section\,\ref{sec:Vcb_BsDs}. 

In Ref.\,\cite{Aaij:2020hsi} the LHCb Collaboration provided also a set of data concerning the differential decay rate $d\Gamma(B_s \to D_s^{(*)} \ell \nu_\ell) / dp_\perp$, where $p_\perp$ is the component of the final $D_s$-meson momentum (i.e., after the strong decay of the $D_s^*$-meson in the case of the $B_s \to D_s^* \ell \nu_\ell$ decays) perpendicular to the flight direction of the $B_s$-meson. The LHCb Collaboration carried out its own estimate of $\vert V_{cb} \vert$ by performing fits of the experimental data on $d\Gamma(B_s \to D_s^{(*)} \ell \nu_\ell)  / dp_\perp$ based on either Caprini-Lellouch-Neubert (CLN)\,\cite{Caprini:1995wq,Caprini:1997mu} or truncated BGL parameterizations\,\cite{Boyd:1995sq, Boyd:1995cf, Boyd:1997kz} of the semileptonic FFs. We make use of the latter fit to reconstruct the experimental data for $d\Gamma(B_s \to D_s^{(*)} \ell \nu_\ell) / dw$ (adopting the updated value of $\vert V_{cb} \vert$ from Ref.\,\cite{LHCb:2021qbv}) in order to get a further determination of $\vert V_{cb} \vert$, as discussed in Section\,\ref{sec:Vcb2nd}. 

In Ref.\,\cite{LHCb:2020hpv} a different LHCb experiment produced the values of the unfolded decay widths for the $B_s \to D_s^* \ell \nu_\ell$ processes integrated in seven $w$-bins and normalized to the total decay rate. These data, together with the total branching fraction from Ref.\,\cite{LHCb:2021qbv}, allow us to determine $\vert V_{cb} \vert$ adopting the bin-per-bin strategy described in Section\,\ref{sec:Vcb1st}.

Finally, in Section\,\ref{sec:comparison} our results for the $B_s \to D_s \ell \nu_\ell$ and $B_s \to D_s^* \ell \nu_\ell$ channels will be compared with other determinations available in the literature and with the most recent  inclusive value of $\vert V_{cb} \vert$.

\subsubsection{$\vert V_{cb} \vert$ from the integrated branching ratios of the $B_s \to D_s^{(*)} \ell \nu_\ell$ decays}
\label{sec:Vcb_BsDs}

The LHCb Collaboration has measured the ratios of the branching fractions of the semileptonic $B_s \to D_s^{(*)} \mu \nu$ decays with respect to the $B \to D^{(*)} \mu \nu$ ones \cite{Aaij:2020hsi}. These measurements read
\bea
    \label{eq:R_exp}
     \frac{\mathcal{B}(B_s \to D_s \mu \nu)}{\mathcal{B}(B \to D \mu \nu)} & = & 1.09 \pm 0.05 \pm 0.06 \pm 0.05 = 1.09 \pm 0.09 ~ , ~ \\[2mm]
    \label{eq:Rstar_exp}
    \frac{\mathcal{B}(B_s \to D_s^{*} \mu \nu)}{\mathcal{B}(B \to D^* \mu \nu)} & = &1.06 \pm 0.05 \pm 0.07 \pm 0.05 = 1.06 \pm 0.10 ~ , ~ 
\eea
where the first error is statistical, the second one is systematic (including the uncertainty related to the choice of the CLN or BGL parameterization) and the third one is due to uncertainties of external inputs used in the measurements. 
Note that the LHCb data indicate that $SU(3)_F$ breaking effects on the branching ratios do not exceed the $\sim 10\%$ level.
Then, the LHCb Collaboration adopted the measured values of $\mathcal{B}(B \to D^{(*)} \mu \nu)$ from PDG\,\cite{ParticleDataGroup:2020ssz} to determine for the first time the branching ratios $\mathcal{B}(B_s \to D_s^{(*)} \mu \nu)$, obtaining
\bea
    \label{eq:BRBsDs_old}
    \mathcal{B}(B_s \to D_s \mu \nu) & = & (2.49 \pm 0.12 \pm 0.14 \pm 0.16) \cdot 10^{-2} = (2.49 \pm 0.24) \cdot 10^{-2} ~ , ~ \\[2mm]
    \label{eq:BRBsDsstar_old}
    \mathcal{B}(B_s \to D_s^* \mu \nu) & = & (5.38 \pm 0.25 \pm 0.46 \pm 0.30) \cdot 10^{-2} = (5.38 \pm 0.60) \cdot 10^{-2} ~ , ~
\eea
where the third error includes also the uncertainty related to the normalization of the branching fractions.
Thanks to an improved determination of the ratio of the $B_s$ and $B$ fragmentation fractions, $f_s / f_d$, the LHCb Collaboration has recently updated\,\cite{LHCb:2021qbv} the above values, obtaining
\bea
    \label{eq:BRBsDs}
    \mathcal{B}(B_s \to D_s \mu \nu) & = & (2.40 \pm 0.22) \cdot 10^{-2} ~ , ~ \\[2mm]
    \label{eq:BRBsDsstar}
    \mathcal{B}(B_s \to D_s^* \mu \nu) & = & (5.19 \pm 0.56) \cdot 10^{-2} ~ , ~
\eea

Using the latest PDG value for the $B_s$-meson lifetime, $\tau_{B_s} = (1.516 \pm 0.006) \cdot 10^{-12} ~ \mbox{s}$\,\cite{ParticleDataGroup:2020ssz}, one has 
\bea
    \label{eq:GammaBsDs_exp}
    \Gamma^{\rm LHCb}(B_s \to D_s \mu \nu) & = & (1.04 \pm 0.10) \cdot 10^{-14} ~ \rm{GeV} ~ , ~ \\[2mm]
    \label{eq:GammaBsDsstar_exp}
    \Gamma^{\rm LHCb}(B_s \to D_s^* \mu \nu) & = & (2.26 \pm 0.24) \cdot 10^{-14} ~ \rm{GeV}  ~ . ~
\eea

Thus, since
\be
    \Gamma(B_s \to D_s^{(*)} \mu \nu) = \int dq^2 \frac{d\Gamma}{dq^2}(B_s \to D_s^{(*)} \mu \nu) ~ , ~ \nonumber
\ee
where the differential decay width $d\Gamma / dq^2$ is given by Eq.\,(\ref{eq:GammaBsDs}) for the $B_s \to D_s \ell \nu_\ell$ decays and Eq.\,(\ref{eq:GammaBsDsstar}) for the $B_s \to D_s^* \ell \nu_\ell$ decays, we can use the DM bands for the FFs given in Figs.\,\ref{fig:FFMMBsDs} and \ref{fig:FFMMBsDsstar} to estimate the theoretical value of the total decay widths modulo $\vert V_{cb} \vert^2$, obtaining
\bea
    \left[ \Gamma(B_s \to D_s \mu \nu) / \vert V_{cb} \vert^2 \right]^{\rm DM} & = & (6.04 \pm 0.23) \cdot 10^{-12}  ~ \rm{GeV} ~ , ~ \\[2mm]
   \left[ \Gamma(B_s \to D_s^* \mu \nu) / \vert V_{cb} \vert^2 \right]^{\rm DM}  & = & (1.39 \pm 0.11) \cdot 10^{-11}  ~ \rm{GeV} ~ . ~
\eea

In this way from Eqs.\,(\ref{eq:GammaBsDs_exp}) -(\ref{eq:GammaBsDsstar_exp}) we get the values
\bea
    \label{eq:Vcb_BsDs_1}
    \vert V_{cb} \vert \cdot 10^3 & = & 41.5 \pm 2.1 ~ \qquad \mbox{from $B_s \to D_s \ell \nu_\ell$ decays} ~ , ~ \\[2mm]
    \label{eq:Vcb_BsDsstar_1}
                                                 & = & 40.3 \pm 2.7 ~ \qquad \mbox{from $B_s \to D_s^* \ell \nu_\ell$ decays} ~ . ~
\eea

\subsubsection{$\vert V_{cb} \vert$ from the differential decay rates of the $B_s \to D_s^{(*)} \ell \nu_\ell$ decays}
\label{sec:Vcb2nd}

In Ref.\,\cite{Aaij:2020hsi} the LHCb Collaboration fitted the $p_{\perp}$ distribution for both the $B_s \to D_s \ell \nu_\ell$ and the $B_s \to D_s^* \ell \nu_\ell$ processes by describing the semileptonic FFs either through a CLN or a truncated BGL parameterizations. The experimental data are not presented explicitly. Instead, the LHCb Collaboration provides the results of their own fits of the experimental data, i.e.~their estimate of $\vert V_{cb} \vert$ (updated in Ref.\,\cite{LHCb:2021qbv}) and of the marginalized values of the parameters entering the CLN or the BGL parameterizations, together with the correlation matrix relating all these quantities to each other.

\begin{figure}[htb!]
\begin{center}
\includegraphics[scale=0.50]{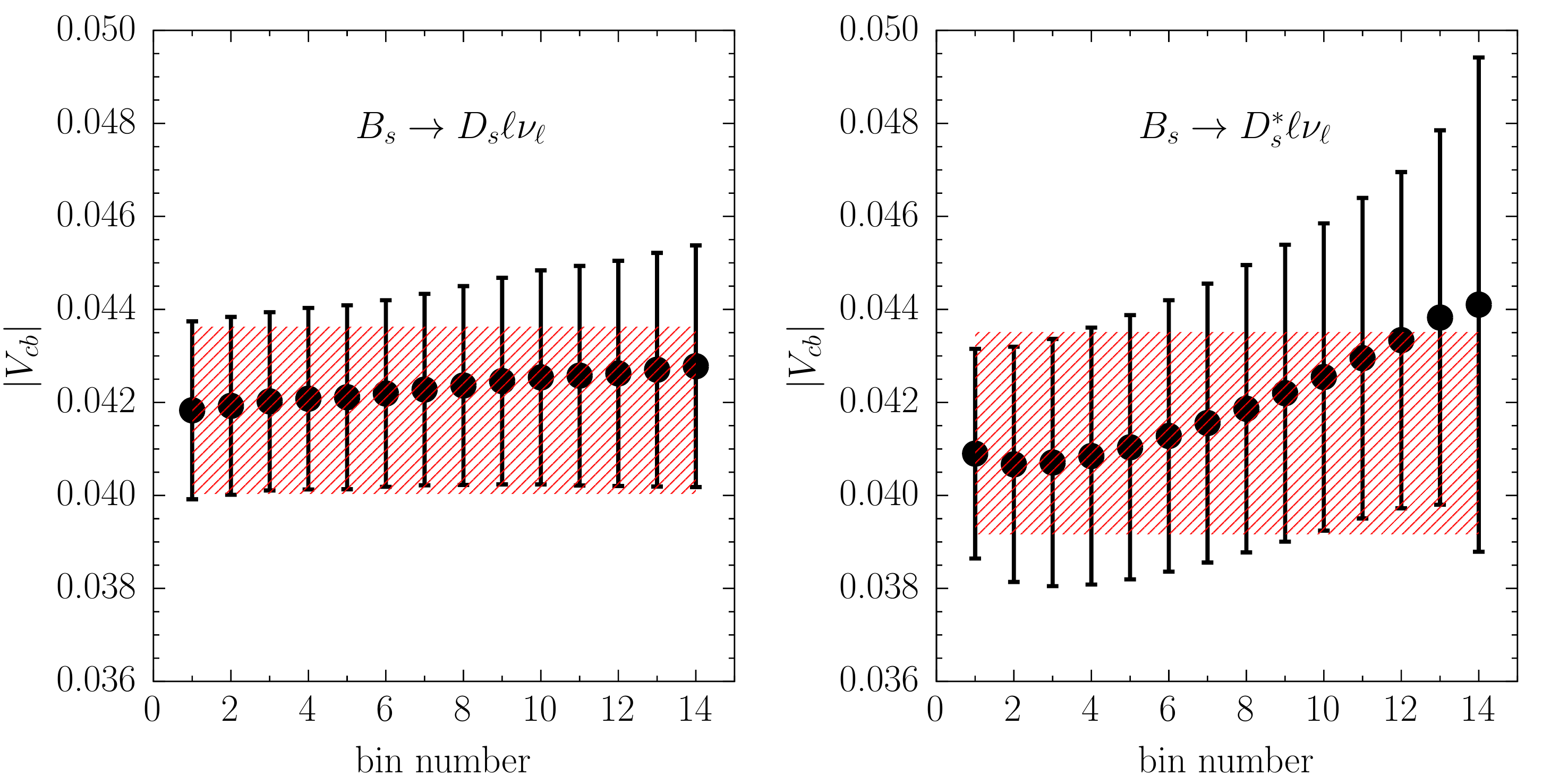}
\vspace{-0.5cm}
\caption{\it \small Bin-per-bin estimates of $\vert V_{cb} \vert$ given by Eq.\,(\ref{eq:Vcb_w}) using the (reconstructed) LHCb data from Refs.\,\cite{Aaij:2020hsi,LHCb:2021qbv} for the $B_s \to D_s \ell \nu_\ell$ (left panel) and $B_s \to D_s^* \ell \nu_\ell$ (right panel) decays. The red bands correspond to the results\,(\ref{eq:Vcb_BsDs_2}) (left panel) and (\ref{eq:Vcb_BsDsstar_2}) (right panel) obtained from the constant fit given in Eqs.\,(\ref{eq:muVcbcorr})-(\ref{eq:sigmaVcbcorr}).}
\label{fig:Vcb_1}
\end{center}
\end{figure}

Using the results of their BGL fit we reconstruct the experimental values of the differential decay width $d\Gamma^{\rm exp}(B_s \to D_s^{(*)} \ell \nu_\ell) / dw$ in a series of single points of the recoil $w$, namely $\{ w_j \} = \{ 1.026, 1.073, 1.121, 1.144, 1.168, 1.215, 1.262, 1.310, 1.357, 1.397, 1.405$, $1.452, 1.499, 1.547\}$ for the $B_s \to D_s \ell \nu_\ell$ decays and $\{ w_j \} = \{ 1.055, 1.109, 1.140, 1.169$, $1.195, 1.221, 1.245, 1.272, 1.295, 1.323$, $1.350, 1.381, 1.425, 1.467\}$ for the $B_s \to D_s^* \ell \nu_\ell$ decays with $j =1, 2, ..., 14$. Note that, following a recommendation from Ref.\,\cite{MD}, we have not changed the correlations of the updated value of $\vert V_{cb} \vert$ from Ref.\,\cite{LHCb:2021qbv} with the BGL coefficients of Ref.\,\cite{Aaij:2020hsi}. Then, using the DM bands for the FFs given in Figs.\,\ref{fig:FFMMBsDs} and \ref{fig:FFMMBsDsstar} we evaluate for each value of $w_j$ the corresponding theoretical expectations modulo $\vert V_{cb} \vert^2$, i.e.~$d\Gamma^{\rm DM}(B_s \to D_s^{(*)} \ell \nu_\ell) / dw$, to get a bin-per-bin estimate of $\vert V_{cb} \vert$, viz.
\be
    \label{eq:Vcb_w}
    \vert V_{cb} \vert_j \equiv \sqrt{\frac{d\Gamma^{\rm exp} / dw_j}{d\Gamma^{\rm DM} / dw_j}} ~ \qquad j = 1, 2, ..., 14 ~ . ~
\ee 
The results for $\vert V_{cb} \vert_j$ are shown in Fig.\,\ref{fig:Vcb_1} as the black dots.

The covariance matrix $C_{ij}$ for the quantities $\vert V_{cb} \vert_j$ can be calculated using (uncorrelated) samples of events for $d\Gamma^{\rm exp} / dw_j$ and $d\Gamma^{\rm DM} / dw_j$ generated according to their respective covariance matrices.
Thus, we determine the value of $\vert V_{cb} \vert$ from a constant fit as
\bea
    \label{eq:muVcbcorr}
    \vert V_{cb} \vert & = & \frac{\sum_{i, j = 1}^{N_{\rm bins}} (\mathbf{C}^{-1})_{ij} \vert V_{cb} \vert_j}{\sum_{i, j = 1}^{N_{\rm bins}} (\mathbf{C}^{-1})_{ij}} ~ , ~ \\[2mm]
    \label{eq:sigmaVcbcorr}
    \sigma^2_{\vert V_{cb} \vert} & = & \frac{1}{\sum_{i, j=1}^{N_{\rm bins}} (\mathbf{C}^{-1})_{ij}} ~ 
\eea
with $N_{\rm bins} = 14$. This procedure leads to the results
\bea
    \label{eq:Vcb_BsDs_2}
     \vert V_{cb} \vert \cdot 10^3 & = & 41.8 \pm 1.8 ~ \qquad \mbox{from $B_s \to D_s \ell \nu_\ell$ decays} ~ , ~ \\[2mm]
    \label{eq:Vcb_BsDsstar_2}
                                                  & = & 41.3 \pm 2.2 ~ \qquad \mbox{from $B_s \to D_s^* \ell \nu_\ell$ decays} ~ , ~
\eea
shown in Fig.\,\ref{fig:Vcb_1} as the red bands.

\subsubsection{$\vert V_{cb} \vert$ from the $B_s \to D_s^* \ell \nu_\ell$ data of Ref.\,\cite{LHCb:2020hpv}}
\label{sec:Vcb1st}

In Ref.\,\cite{LHCb:2020hpv} a different LHCb experiment has provided the values of the ratios
\be
    \label{eq:ratiosj}
    \Delta r_j \equiv \frac{\Delta \Gamma_j(B_s \to D_s^* \mu \nu)}{\Gamma(B_s \to D_s^* \mu \nu)} ~ \qquad \mbox{j = 1, 2, ..., 7}
\ee
between the decay rate $\Delta \Gamma_j(B_s \to D_s^* \mu \nu)$ integrated in each of seven $w$-bins with the total decay rate $\Gamma(B_s \to D_s^* \mu \nu)$.
The experimental data are collected in Table\,\ref{tab:DiffRateComparison} and compared with the corresponding predictions of the DM method based on the FFs of Fig.\,\ref{fig:FFMMBsDsstar} obtained starting from the lattice inputs of Table\,\ref{tab:LQCDBsDsstar}.
Our DM results turn out to be consistent with the corresponding ones calculated by HPQCD Collaboration in Ref.\,\cite{Harrison:2021tol}.

\begin{table}[htb!]
\renewcommand{\arraystretch}{1.2}
\begin{center}
\begin{adjustbox}{max width=\textwidth}
\begin{tabular}{|c||c|c|c|c|c|c|c|}
\hline
~ $j$ ~ & 1 & 2 & 3 & 4 & 5 & 6 & 7\\
\hline
~ $w$-bin ~ & 1.000 - 1.1087 & 1.1087 - 1.1688 & 1.1688 - 1.2212  & 1.2212 - 1.2717 & 1.2717 - 1.3226 & 1.3226 - 1.3814 & 1.3814 - 1.4667\\
\hline
~ $\Delta w_j$ ~ & 0.1087 & 0.0601 & 0.0524 & 0.0505 & 0.0509 & 0.0588 & 0.0853\\
\hline \hline
~ $\Delta r_j^{\rm LHCb}$ ~ & 0.183(12) & 0.144(8) & 0.148(8) & 0.128(8) & 0.117(7) & 0.122(6) & 0.158(9)\\
\hline \hline
~ $\Delta r_j^{\rm DM}$ ~ & 0.1942(82) & 0.1534(45) & 0.1377(28)& 0.1289(18) & 0.1212(20) & 0.1241(40) & 0.1405(110) \\
\hline \hline
\end{tabular}
\end{adjustbox}
\vspace{0.10cm}
\caption{\it \small Values of the ratios $\Delta r_j$ given in Eq.\,(\ref{eq:ratiosj}) for each of the seven experimental $w$-bins of Ref.\,\cite{LHCb:2020hpv}. The w-bins and their widths $\Delta w_j$ are given in the second and third rows, respectively.  The forth row collects the experimental data from Ref.\,\cite{LHCb:2020hpv}. The last row corresponds to the theoretical results obtained using the FFs shown in Fig.\,\ref{fig:FFMMBsDsstar} and determined by the DM method starting from the lattice inputs of Table\,\ref{tab:LQCDBsDsstar}.}
\label{tab:DiffRateComparison}
\end{center}
\renewcommand{\arraystretch}{1.0}
\end{table}

\begin{figure}[htb!]
\begin{center}
\includegraphics[scale=0.50]{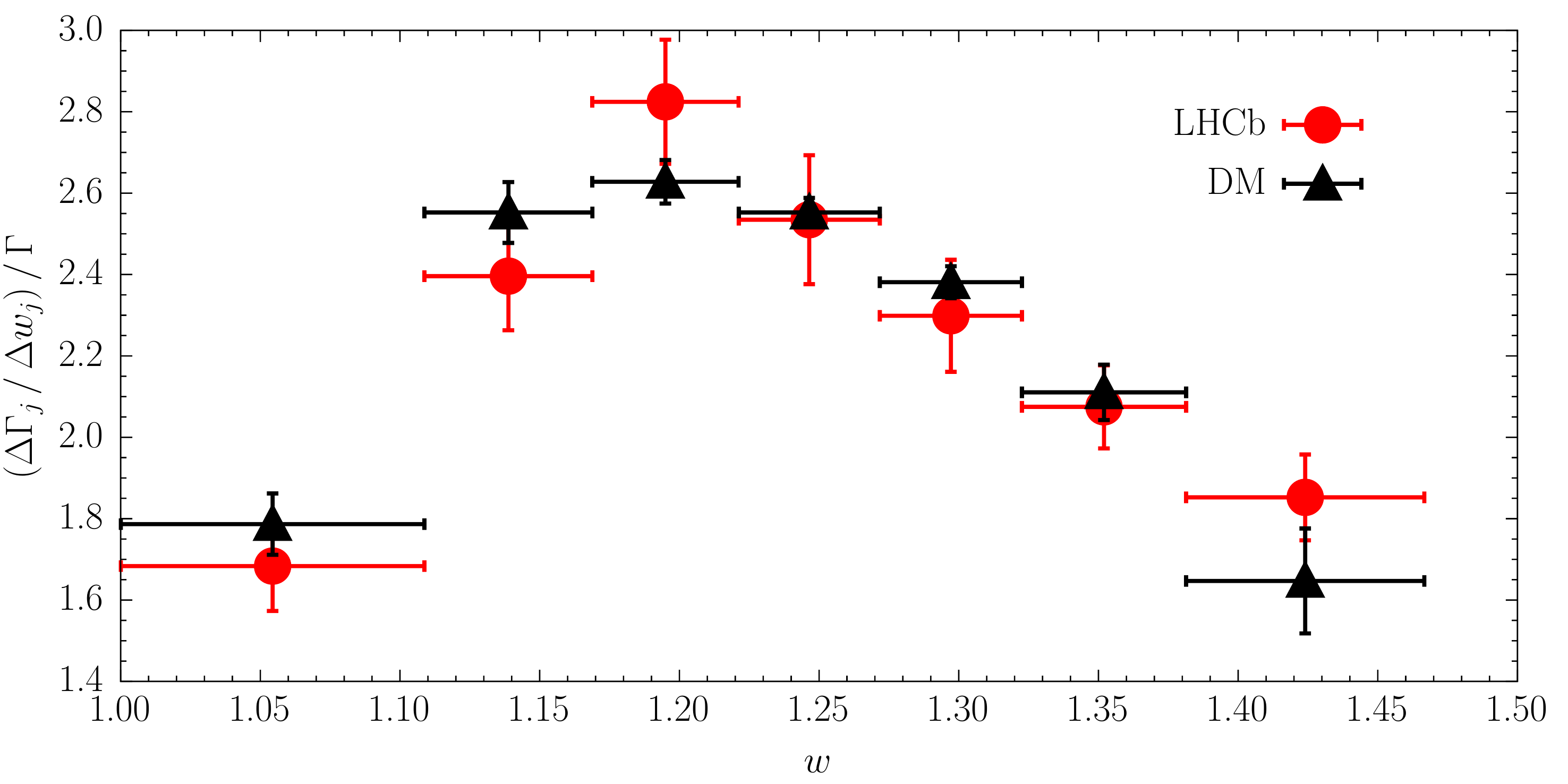}
\vspace{-0.5cm}
\caption{\it \small The differential decay rate $\Delta r_j / \Delta w_j = (\Delta \Gamma_j / \Delta w_j ) / \Gamma$, integrated for each of the seven experimental $w$-bins of Ref.\,\cite{LHCb:2020hpv} and normalized by the total decay rate, versus the recoil variable $w$. The red dots represent the LHCb experimental data of Ref.\,\cite{LHCb:2020hpv}, while the black triangles are the predictions based on our DM approach starting from the lattice inputs of Table\,\ref{tab:LQCDBsDsstar}.}
\label{fig:ratio}
\end{center}
\end{figure}

In Fig.\,\ref{fig:ratio} the differential decay rates $\Delta r_j / \Delta w_j = (\Delta \Gamma_j / \Delta w_j ) / \Gamma$ are compared for each of the seven $w$-bins with the corresponding experimental data of Ref.\,\cite{LHCb:2020hpv}.
It can be seen that the shape of the theoretical predictions is consistent with the one of the experimental data within $\approx 1$ standard deviation.

Using the experimental value\,(\ref{eq:GammaBsDsstar_exp}) for the total decay rate $\Gamma(B_s \to D_s^* \mu \nu)$ we can compute the experimental values of the (partially) integrated decay rate $\Delta \Gamma_j$ for each $w$-bin as
\be
    \label{eq:Gamma_w}
    \Delta \Gamma_j^{exp} = \Delta r_j^{\rm LHCb} \cdot \Gamma^{\rm LHCb}(B_s \to D_s^* \mu \nu) ~ . ~
\ee
The covariance matrix $\Gamma_{ij}^{exp}$ for the  decay rates $\Delta \Gamma_j^{exp}$ is evaluated by considering a sample of events for the ratios $\Delta r_j^{\rm LHCb}$ generated according to the experimental covariance matrix $R_{ij}^{\rm LHCb}$ provided in Ref.\,\cite{LHCb:2020hpv} and a Gaussian distribution for $\Gamma^{\rm LHCb}(B_s \to D_s^* \mu \nu)$ with mean value $\overline{\Gamma} = 2.26 \cdot 10^{-14}$ GeV and standard deviation $\sigma_{\overline{\Gamma}} = 0.24 \cdot 10^{-14}$ GeV. The latter distribution is uncorrelated with those of the ratios $\Delta r_j^{\rm LHCb}$, since it comes from a different LHCb experiment.
A simple calculation yields
\be
    \label{eq:Gammaij_exp}
    \Gamma_{ij}^{exp} = R_{ij}^{\rm LHCb} \left[ \overline{\Gamma}^2 + \sigma_{\overline{\Gamma}}^2 \right] + 
                                      \Delta r_i^{\rm LHCb} \Delta r_j^{\rm LHCb} \sigma_{\overline{\Gamma}}^2 ~ . ~
\ee

Notice that, since the sum of the ratio $\Delta r_j^{\rm LHCb}$ over the seven $w$-bins is equal to unity by construction, the covariance matrix $R_{ij}^{\rm LHCb}$ must have a null eigenvalue, so that the number of independent bins is six. This does not occur for the original covariance matrix provided in Ref.\,\cite{LHCb:2020hpv}. 
Thus, we generate a sample of events for the seven $w$-bins using the multivariate Gaussian distribution corresponding to the original covariance matrix of the ratios. Then, for each event we normalize the ratios by their sum over the bins and recalculate the covariance matrix, which has now properly a null eigenvalue. 
In what follows we make use of the corrected covariance matrix, though the numerical impact of the correction on the determination of $\vert V_{cb} \vert$ turns out to be negligible.

\begin{figure}[htb!]
\begin{center}
\includegraphics[scale=0.50]{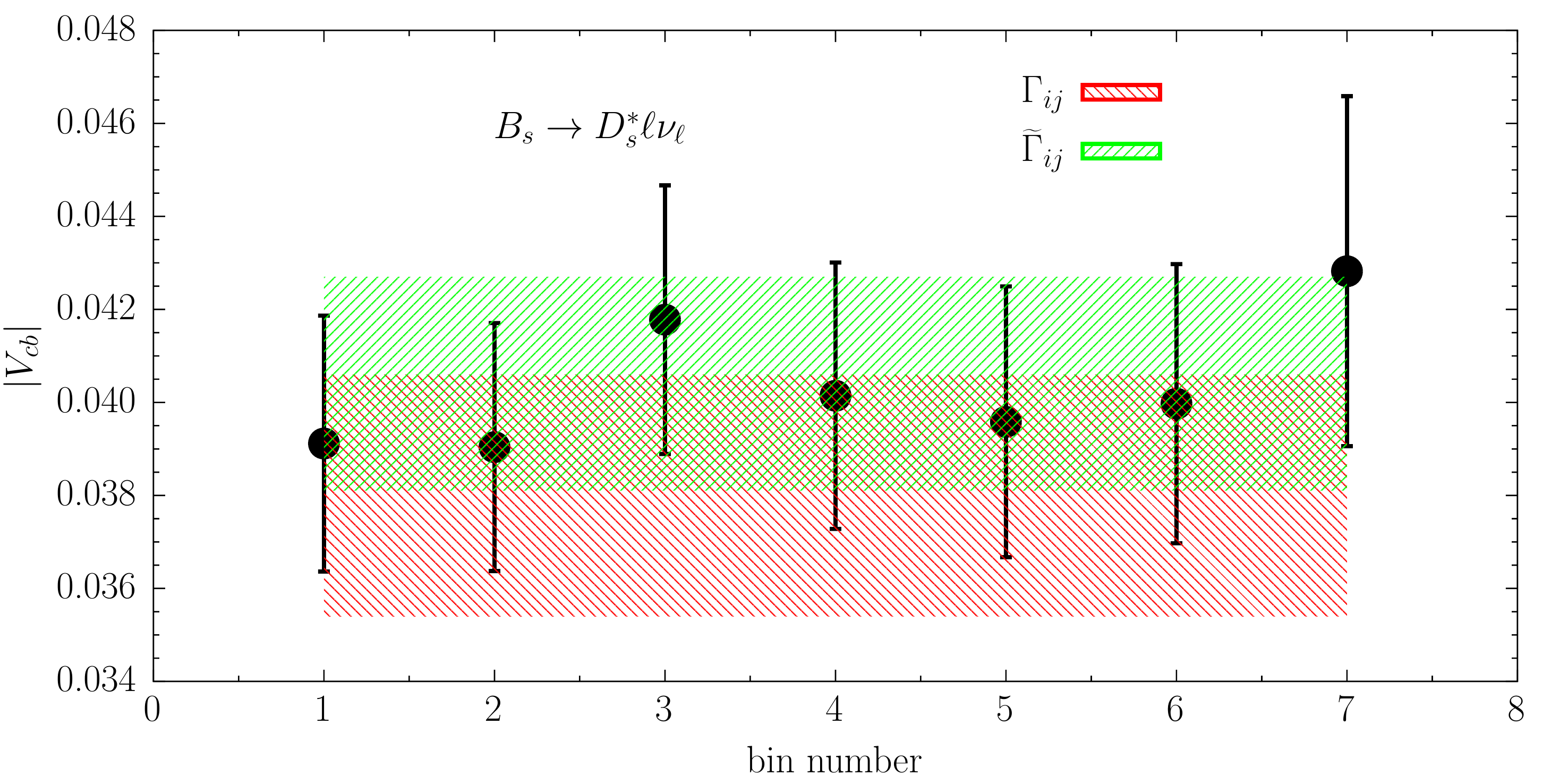}
\vspace{-0.5cm}
\caption{\it \small Bin-per-bin estimates of $\vert V_{cb} \vert$ given by Eq.\,(\ref{eq:Vcbj}) using the LHCb data of Ref.\,\cite{LHCb:2020hpv}. The red and green bands correspond respectively to the values\,(\ref{eq:Vcb_BsDsstar_3}) and\,(\ref{eq:Vcb_BsDsstar_3der}). They are obtained by the constant fit given in Eqs.\,(\ref{eq:muVcbcorr})-(\ref{eq:sigmaVcbcorr}) using for the experimental data $\Delta \Gamma_j^{exp}$ the covariance matrices\,(\ref{eq:Gammaij_exp}) (red bands) and \,(\ref{eq:Gammaij_Der}) (green bands).}
\label{fig:Vcb_2}
\end{center}
\end{figure}

Using the DM bands of the FFs we now evaluate the theoretical predictions $\Delta \Gamma_j^{DM}$ (and the corresponding covariance matrix $\Gamma_{ij}^{DM}$) that can be compared with the experimental ones\,(\ref{eq:Gamma_w}) to obtain the value of $\vert V_{cb} \vert$ for each of the seven $w$-bins, namely
\be
    \label{eq:Vcbj}
    \vert V_{cb} \vert_j \equiv \sqrt{\frac{\Delta \Gamma_j^{exp}}{\Delta \Gamma_j^{DM}}} ~ . ~
\ee
The results for $\vert V_{cb} \vert_j $ are shown in Fig.\,\ref{fig:Vcb_2} as the black dots.
The covariance matrix $C_{ij}$ for the quantities $\vert V_{cb} \vert_j$ can be calculated using (uncorrelated) samples of events for $\Delta \Gamma_j^{exp}$ and $\Delta \Gamma_j^{DM}$ generated according to their respective covariance matrices $\Gamma_{ij}^{exp} $ and $\Gamma_{ij}^{DM}$.

Finally, we determine the value of $\vert V_{cb} \vert$ from the constant fit \,(\ref{eq:muVcbcorr})-(\ref{eq:sigmaVcbcorr}) with $N_{\rm bins} = 7$. 
This procedure leads to the result
\be
    \label{eq:Vcb_BsDsstar_3}
    \vert V_{cb} \vert  \cdot 10^3 = 38.0 \pm 2.6 ~ , ~
\ee
shown in Fig.\,\ref{fig:Vcb_2} as the red band. 
It can be seen that the central value of Eq.\,(\ref{eq:Vcb_BsDsstar_3}) lies well below the bin-per-bin data. This problem is well-known in literature\,\cite{DAgostini:1993arp} and we have already addressed it in Ref.\,\cite{Martinelli:2021onb}. The issue is related to the fact that best fits to data which are affected by an overall normalization uncertainty (like in our case $\sigma_{\overline{\Gamma}}$ for the quantities $ \Delta \Gamma_j^{exp}$) have the tendency to produce curves lower than expected if the covariance matrix of the data points is used in the definition of the $\chi^2$-variable. In particular, in Ref.\,\cite{DAgostini:1993arp} it was shown that in the case of a fit to a constant, like Eqs.\,(\ref{eq:muVcbcorr})-(\ref{eq:sigmaVcbcorr}), a negative bias can be obtained, the absolute size of which is proportional to the number of degrees of freedom, to the square of the normalization uncertainty (i.e., to $\sigma_{\overline{\Gamma}}^2$) and to the differences between individual data points (i.e., in our case to the differences among the ratios $\Delta r_j^{\rm LHCb}$).
In other words the observed negative bias is due to the term $\Delta r_i^{\rm LHCb} \Delta r_j^{\rm LHCb} \sigma_{\overline{\Gamma}}^2$ appearing in the r.h.s.~of Eq.\,(\ref{eq:Gammaij_exp}) and it is driven by $\sigma_{\overline{\Gamma}} \neq 0$ and/or $\Delta r_i^{\rm LHCb} \neq \Delta r_j^{\rm LHCb}$.
The offending term $\Delta r_i^{\rm LHCb} \Delta r_j^{\rm LHCb} \sigma_{\overline{\Gamma}}^2$, however, is necessary to guarantee: i) the normalization property $\sum_{i,j =1}^{N_{\rm bins}} \Gamma_{ij}^{exp} = \sigma_{\overline{\Gamma}}^2$, which follows from the fact that $\sum_{j =1}^{N_{\rm bins}} \Delta r_j^{\rm LHCb} = 1$ (or equivalently $\sum_{j =1}^{N_{\rm bins}} \Delta \Gamma_j^{exp} = \overline{\Gamma}$)\footnote{Notice that also the relation $\sum_{i,j =1}^{N_{\rm bins}} R_{ij}^{\rm LHCb} = 0$ holds as well, as we have explicitly checked.}, and ii) the existence of an inverse of $\Gamma_{ij}$, because the (corrected) matrix $R_{ij}^{\rm LHCb}$ has a null eigenvalue.

We now follow a suggestion described in Ref.\,\cite{DAgostini:1993arp} for solving the problem of the negative bias. It corresponds to consider in Eq.\,(\ref{eq:Gammaij_exp}) the limiting case in which all the ratios $\Delta r_i^{\rm LHCb}$ are equal to each other, and therefore equal to $1 / N_{bins}$. In this way the constant fit to the data points is not plagued any more by a negative bias. In other words a new estimate of the covariance matrix is given by
\be
    \label{eq:Gammaij_Der}
    \widetilde{\Gamma}_{ij}^{exp} = R_{ij}^{\rm LHCb} \left[ \overline{\Gamma}^2 + \sigma_{\overline{\Gamma}}^2 \right] + 
                                                      \frac{1}{N_{\rm bins}^2} \sigma_{\overline{\Gamma}}^2 ~ , ~
\ee
which has an inverse and still fulfills the normalization property $\sum_{i,j =1}^{N_{\rm bins}} \widetilde{\Gamma}_{ij}^{exp} = \sigma_{\overline{\Gamma}}^2$.
Using the modified covariance matrix $\widetilde{\Gamma}_{ij}^{exp}$ the constant fit\,(\ref{eq:muVcbcorr})-(\ref{eq:sigmaVcbcorr}) yields the result
\be
    \label{eq:Vcb_BsDsstar_3der}
    |V_{cb}| \cdot 10^3 = 40.4 \pm 2.3 ~ , ~
\ee
shown in Fig.\,\ref{fig:Vcb_2} as the green band. The previous negative bias is now removed.

\subsubsection{Comparison with the inclusive and the other exclusive estimates of $\vert V_{cb} \vert$}
\label{sec:comparison}

After averaging the results\,(\ref{eq:Vcb_BsDs_1}) and (\ref{eq:Vcb_BsDs_2}) for the semileptonic $B_s \to D_s \ell \nu_\ell$ decays and the results\,(\ref{eq:Vcb_BsDsstar_1}), (\ref{eq:Vcb_BsDsstar_2}) and (\ref{eq:Vcb_BsDsstar_3der}) for the semileptonic $B_s \to D_s^* \ell \nu_\ell$ decays\footnote{Since the individual determinations are not independent we follow the procedure already adopted in Refs.\,\cite{Martinelli:2021myh,Martinelli:2022tte}: starting from $N$ computations with mean values $x_k$ and uncertainties $\sigma_k$ ($k=1,\cdots,N$), the \emph{combined} average $x$ and uncertainty $\sigma_x$ are given by (see Ref.\,\cite{EuropeanTwistedMass:2014osg})
\be
    x = \sum_{k=1}^N \omega_k x_k ~ , ~ \qquad
    \sigma_x^2 = \sum_{k=1}^N \omega_k \sigma_k^2 + \sum_{k=1}^N \omega_k (x_k - x)^2 ~ , ~
\ee
where $\omega_k$ represents the weight associated to the $k$-th determination. We assume $\omega_k = (1/ \sigma_k^2) / \sum_{j=1}^N (1/ \sigma_j^2)$.} we get
\bea
    \label{eq:Vcb_BsDs}
     \vert V_{cb} \vert^{\rm DM} \cdot 10^3 & = & 41.7 \pm 1.9 ~ \qquad \mbox{from $B_s \to D_s \ell \nu_\ell$ decays} ~ , ~ \\[2mm]
    \label{eq:Vcb_BsDsstar}
                                                                  & = & 40.7 \pm 2.4 ~ \qquad \mbox{from $B_s \to D_s^* \ell \nu_\ell$ decays} ~ . ~
\eea

By combining the two above results in a weighted average our final estimate of $\vert V_{cb} \vert$ is given by
\be
    \label{eq:Vcb_DM_Bs}
    \vert V_{cb} \vert^{\rm DM} \cdot 10^3 = 41.3 \pm 1.5 ~ \qquad \mbox{from $B_s \to D_s^{(*)} \ell \nu_\ell$ decays} ~ . ~
\ee

Our finding\,(\ref{eq:Vcb_DM_Bs}) agrees with 
the estimate made by the LHCb Collaboration in Ref.\,\cite{LHCb:2021qbv} (using their truncated BGL fit)
\be
    \label{eq:Vcb_LHCb}
    \vert V_{cb} \vert^{\rm LHCb} \cdot 10^3 = 41.7 \pm 0.8 \pm 0.9 \pm 1.1 = 41.7 \pm 1.6 ~ \qquad \mbox{from $B_s \to D_s^{(*)} \ell \nu_\ell$ decays} ~ , ~
\ee
where the first uncertainty is statistical, the second one systematic and the third one due to the limited knowledge of some external inputs\footnote{We do not make a comparison with the result of Ref.\,\cite{Harrison:2021tol}, since in our work we use the different updated value of  $\vert V_{cb} \vert$ from Ref.\,\cite{LHCb:2021qbv}.}.

As already mentioned in the Introduction, we applied the DM method to determine $\vert V_{cb} \vert$ from the semileptonic $B \to D^{(*)} \ell \nu_\ell$ decays in Refs.\,\cite{Martinelli:2021onb, Martinelli:2021myh}, obtaining the results given in Eqs.\,(\ref{eq:Vcb_BD})-(\ref{eq:Vcb_BDstar}).
Their weighted average reads
\be
    \label{eq:Vcb_DM_B}
    \vert V_{cb} \vert^{\rm DM} \cdot 10^3 = 41.1 \pm 1.0 ~ \qquad \mbox{from $B \to D^{(*)} \ell \nu_\ell$ decays} ~ , ~
\ee
which is compatible with our result\,(\ref{eq:Vcb_DM_Bs}) from $B_s \to D_s^{(*)} \ell \nu_\ell$ decays.
The final (weighted) average between the DM results\,(\ref{eq:Vcb_DM_Bs}) and (\ref{eq:Vcb_DM_B}) yields\footnote{A correlation between the results \,(\ref{eq:Vcb_DM_Bs}) and (\ref{eq:Vcb_DM_B}) may be generated by: i) the LHCb's use of the PDG branching fractions of the $B \to D^{(*)} \ell \nu_\ell$ processes to obtain those of the $B_s \to D_s^{(*)} \ell \nu_\ell$ decays, ii) the use of the same gauge ensembles for the lattice FFs of different processes, and  iii) the use of the same values of the susceptibilities of the $b \to c$ quark transitions for the various FFs. Information on item ii) is not currently available, while in the cases i) and iii) we have estimated a correlation coefficient of the order of ${\cal{O}}(0.1)$, which has a negligible impact on our final average\,(\ref{eq:Vcb_DM}).}
\be
    \label{eq:Vcb_DM}
    \vert V_{cb} \vert^{\rm DM} \cdot 10^3 = 41.2 \pm 0.8 ~ \qquad \mbox{from $B_{(s)} \to D_{(s)}^{(*)} \ell \nu_\ell$ decays} ~ . ~
\ee
From the latest FLAG review\,\cite{FlavourLatticeAveragingGroupFLAG:2021npn} one has
\be
    \label{eq:Vcb_FLAG}
    \vert V_{cb} \vert^{\rm FLAG} \cdot 10^3 = 39.48 \pm 0.68 ~ \qquad \mbox{from $B \to D^{(*)} \ell \nu_\ell$ decays} ~ , ~
\end{equation}
which is $\simeq 1.6 \sigma$ below our result\,(\ref{eq:Vcb_DM}). Note that the uncertainties in Eqs.\,\,(\ref{eq:Vcb_DM_B})-(\ref{eq:Vcb_FLAG}) are comparable, though the DM approach does not use experimental data to constrain the shape of the hadronic FFs.

All the results\,(\ref{eq:Vcb_DM_Bs})-(\ref{eq:Vcb_DM}) are in agreement with the most recent inclusive determination of $\vert V_{cb} \vert$, which reads $\vert V_{cb} \vert_{\rm{incl}} \cdot 10^3 = 42.16 \pm 0.50$\,\cite{Bordone:2021oof}, at the $1 \sigma$ level.

\subsection{Lepton Flavour Universality and polarization observables}
\label{sec:polarization}

We now compute the theoretical values of the $\tau / \ell$ ratios $R(D_s^{(*)})$ (where $\ell$ is an electron or a muon), the $\tau$-polarization $P_\tau(D_s^*)$ and the $D_s^*$ longitudinal polarization $F_L(D_s^*)$ using the DM bands of the FFs of Figs.\,\ref{fig:FFMMBsDs} and \ref{fig:FFMMBsDsstar}.

The ratios $R(D_s^{(*)})$ are powerful tests of LFU and they are defined as the $\tau/\ell$ ratios of the corresponding total decay rates, namely
\be
    R(D_s^{(*)}) \equiv \frac{\Gamma(B_s \to D_s^{(*)} \tau \nu_{\tau})}{\Gamma(B_s \to D_s^{(*)} \ell \nu)},
\ee
where $\ell$ is a light lepton. Also the quantities $P_{\tau}^s$ and $F_L^s$ depend only on the shape of the semileptonic FFs (see Ref.\,\cite{Ivanov:2016qtw} for their general definitions valid both within the SM and beyond).

Following Refs.\,\cite{Martinelli:2021onb, Martinelli:2021myh} we compute bootstrap values of $R(D_s^{(*)})$, $P_{\tau}^s$ and $F_L^s$ by using the events extracted for the FFs after the implementation of the DM method. We then fit each of the histograms of the resulting events with a normal distribution, in order to obtain a final expectation value and a final uncertainty for each physical quantity. Our results are
\bea
    \label{eq:RDs&RDsstar}
    R(D_s) & = & 0.299 ~ (5) ~ , ~ \qquad \qquad R(D_s^*) = 0.250 ~ (6) ~ , ~ \\
    \label{eq:Ptaus&FLs}
    P_\tau(D_s^*) & = & -0.520 ~ (12) ~ , ~ \qquad ~ F_L(D_s^*) = 0.440 ~ (16) ~ , ~
\eea
which can be compared with the corresponding results obtained in Refs.\,\cite{Martinelli:2021onb, Martinelli:2021myh} in the case of the semileptonic $B \to D^{(*)} \ell \nu_\ell$ decays, namely $R(D) = 0.296~(8)$, $R(D^*) = 0.275~(8)$, $P_\tau(D^*) = -0.529~(7)$ and $F_L(D^*) = 0.414~(12)$.
It follows that $SU(3)_F$ breaking effects are negligible for all the above quantities except $R(D^*)$ and $R(D_s^*)$, which differ by $\approx 10\%$.
Such a difference is a consequence of the different shapes of the semileptonic FFs entering the $B_{(s)} \to D_{(s)}^* \ell \nu_\ell$ decays, as it will be discussed in the next Section.

A few improved versions of the LFU ratios in semileptonic decays of pseudoscalar mesons into vector ones have been proposed recently~\cite{Isidori:2020eyd} in order to minimize the theoretical FF uncertainties. In the case of the semileptonic $B_s \to D_s^*$ decays we consider the following two definitions
\bea
    R^{cut}(D_s^*) & \equiv & \frac{\int_{m_\tau^2}^{q_{max}^2} dq^2 ~ \frac{d\Gamma}{dq^2}(B_s \to D_s^* \tau \nu_\tau)}
                                              {\int_{m_\tau^2}^{q_{max}^2} dq^2 ~ \frac{d\Gamma}{dq^2}(B_s \to D_s^* \ell \nu_\ell)} ~ , ~ \\[2mm]
    R^{opt}(D_s^*) & \equiv & \frac{\int_{m_\tau^2}^{q_{max}^2} dq^2 \frac{d\Gamma}{dq^2}(B_s \to D_s^* \tau \nu_\tau)}
                                              {\int_{m_\tau^2}^{q_{max}^2} dq^2 \left[ \frac{\omega_\tau(q^2)}{\omega_\ell(q^2)} \right] \frac{d\Gamma}{dq^2}(B_s \to D_s^* \ell \nu_\ell)} ~ , ~                              
\eea
where $\omega_\ell(q^2) = (1 - m_\ell^2 / q^2)^2 (1 + m_\ell^2 / 2q^2)$. Using the DM bands of the FFs we get $R^{cut}(D_s^*) = 0.332 \pm 0.003$ and $R^{opt}(D_s^*) = 1.082 \pm 0.008$, which exhibit uncertainties equal to $\simeq 0.9 \%$ and $\simeq 0.7 \%$, respectively.

We remind that our DM estimates of all the LFU ratios come from an average over the $\ell = e$ and $\ell = \mu$ cases.

\begin{figure}[htb!]
\begin{center}
\includegraphics[scale=0.55]{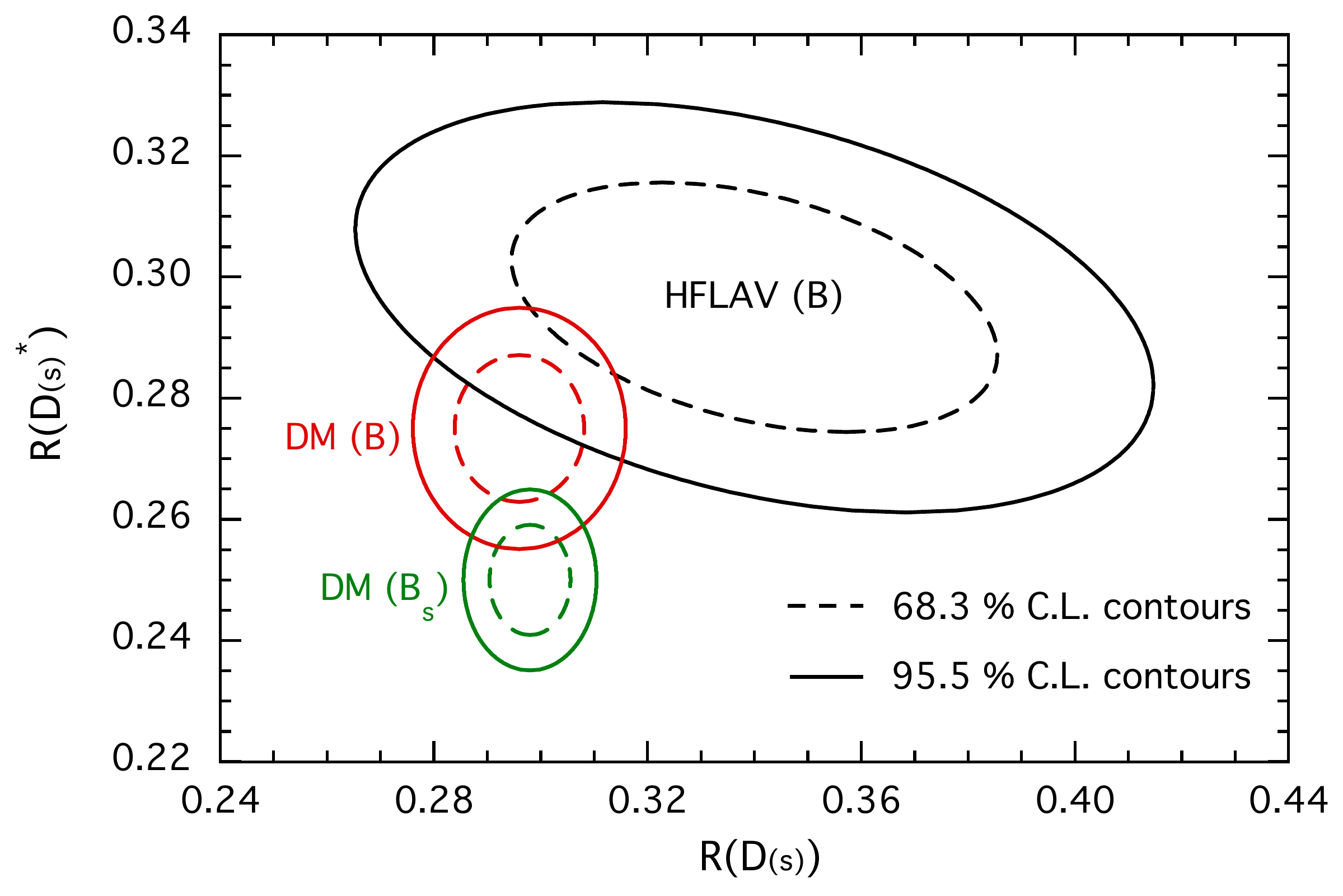}
\vspace{-0.5cm}
\caption{\it \small The correlation plot for $R(D_{(s)})$ and $R(D_{(s)}^*)$. The black area represents the average of all the experimental measurements of $R(D)$ and $R(D^*)$ from Ref.\,\cite{Amhis:2016xyh}. The red and the green regions are the DM predictions for $R(D_{(s)})$ and $R(D_{(s)}^*)$ obtained in Refs.\,\cite{Martinelli:2021onb, Martinelli:2021myh} and in this work, respectively.}
\label{fig:RDRDstar}
\end{center}
\end{figure}

We close this Section by comparing in Fig.\,\ref{fig:RDRDstar} the latest HFLAV averages\,\cite{Amhis:2016xyh} of the experimental measurements of the ratios $R(D)$ and $R(D^*)$ with our DM theoretical expectations for $R(D_{(s)})$ and $R(D_{(s)}^*)$.

\section{Comparison between the $B_s \to D_s^{(*)}$ and $B \to D^{(*)}$ Form Factors}
\label{sec:SU3_F}

In this Section we compare the DM bands of the hadronic FFs entering the $B \to D^{(*)} \ell \nu_\ell$ and  $B_s \to D_s^{(*)} \ell \nu_\ell$ decays obtained in Refs.\,\cite{Martinelli:2021onb, Martinelli:2021myh} and in this work, respectively\footnote{The impact of $SU(3)_F$ breaking effects on the semileptonic FFs has been investigated in Ref.\,\cite{Kobach:2019kfb} using unitarity and analiticity, and also in Ref.\,\cite{Bordone:2019guc} using theoretical results obtained from the Light Cone Sum Rule approach, but mixed with the experimental data to constrain the shape of the FFs.}. In the former case the LQCD inputs for the DM method come from the results of the FNAL/MILC Collaboration\,\cite{MILC:2015uhg, FermilabLattice:2021cdg}. The non-perturbative values of the relevant susceptibilities for the $b \to c$ transitions are the same for all the decays considered, i.e.~they are given by Eqs.\,(\ref{eq:bound0+})-(\ref{eq:bound1+}).

\begin{figure}[htb!]
\begin{center}
\includegraphics[scale=0.50]{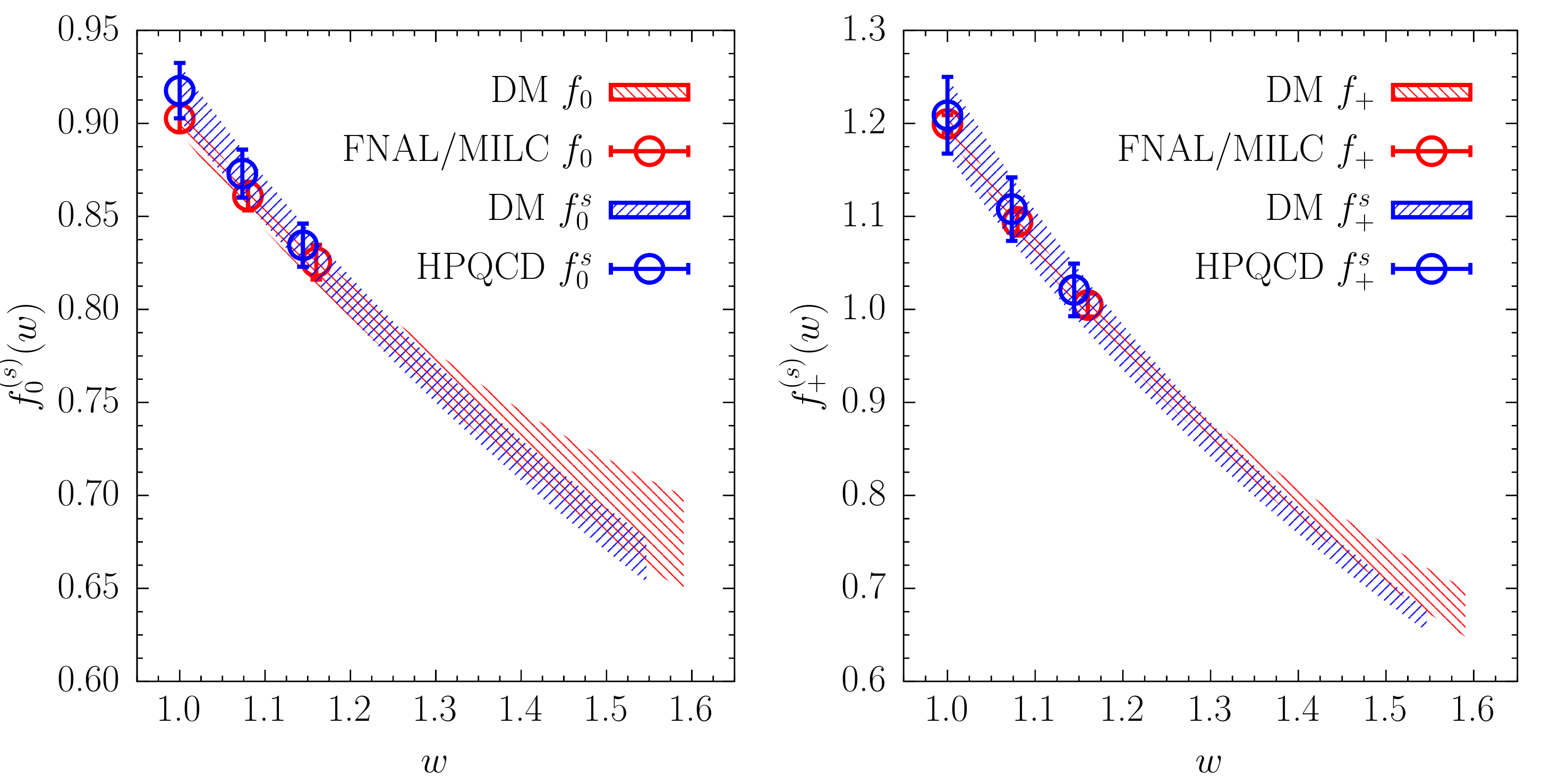}
\vspace{-0.5cm}
\caption{\it \small The DM bands of the scalar and vector FFs entering the semileptonic $B_{(s)} \to D_{(s)} \ell \nu_\ell$ decays, i.e.~$f_0^{(s)}(w)$ (left panel) and $f_+^{(s)}(w)$ (right panel) versus the recoil variable $w$. The red and blue bands correspond to the $B \to D$ and $B_s \to D_s$ transitions, evaluated in Ref.\,\cite{Martinelli:2021onb} and in this work, respectively. Correspondingly the red and blue circles represent the LQCD results used as inputs in the DM method, coming respectively from Refs.\,\cite{MILC:2015uhg} and \cite{McLean:2019qcx}.}
\label{fig:FFMMBD}
\end{center}
\end{figure}

The DM bands for the two sets of scalar and vector FFs of the $B_{(s)} \to D_{(s)} \ell \nu$ transitions are shown in Fig.\,\ref{fig:FFMMBD} and turn out to be compatible with quite small differences in their slopes. Thus, for the transitions to a final pseudoscalar meson we observe small $SU(3)_F$ breaking effects. 

\begin{figure}[htb!]
\begin{center}
\includegraphics[scale=0.50]{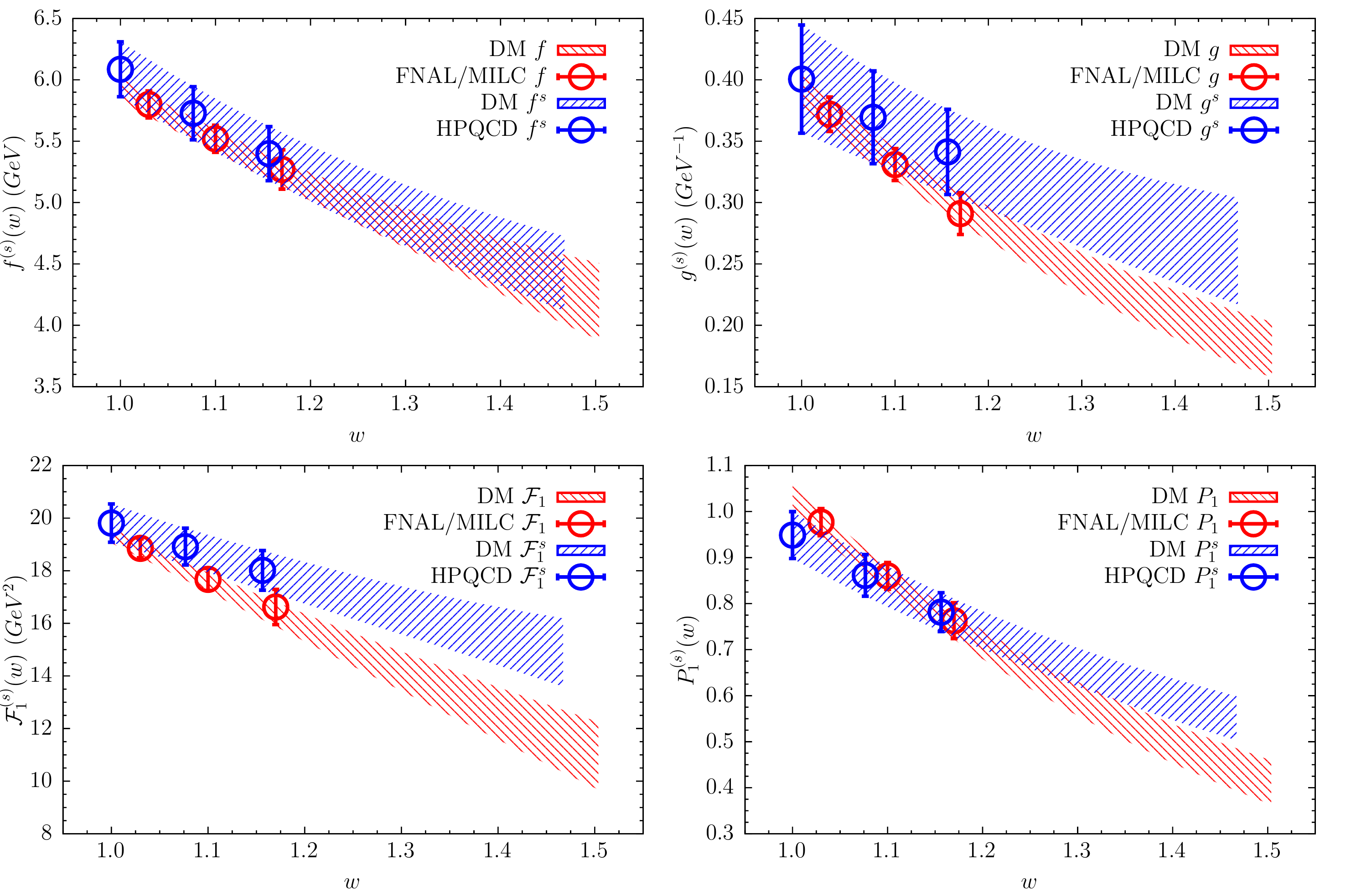}
\vspace{-0.5cm}
\caption{\it \small The DM bands of the four FFs entering the semileptonic $B_{(s)} \to D_{(s)}^* \ell \nu$ decays, i.e.~$f^{(s)}(w)$, $g^{(s)}(w)$, $\mathcal{F}^{(s)}_1(w)$ and $P_1^{(s)}(w)$ versus the recoil variable $w$. The red and blue bands correspond to the $B \to D^*$ and $B_s \to D_s^*$ transitions, evaluated in Ref.\,\cite{Martinelli:2021myh} and in this work, respectively. Correspondingly the red and blue circles represent the LQCD results used as inputs in the DM method, coming respectively from Refs.\,\cite{FermilabLattice:2021cdg} and \cite{Harrison:2021tol}.}
\label{fig:FFMMBDstar}
\end{center}
\end{figure}

The DM bands for the two sets of FFs of the $B_{(s)} \to D_{(s)}^{*} \ell \nu$ transitions are collected in Fig.\,\ref{fig:FFMMBDstar}. In this case the FFs $f$ and $g$ exhibit small $SU(3)_F$ breaking effects, while the shapes for the other two FFs are remarkably different. More quantitatively, the ratio of the mean values of the FFs $\mathcal{F}_1^s(w)$ and $\mathcal{F}_1(w)$ at $ w = w_{max}^*$ (the maximum recoil for the $B_s \to D_s^* \ell \nu_\ell$ decays) shows a $\sim 30\%$ deviation from unity. Assuming the two FFs as uncorrelated (such a correlation is not known presently), we get the na\"ive estimate $\mathcal{F}_1^s(w_{max}^*)/\mathcal{F}_1(w_{max}^*) = 1.29 \pm 0.17$. A similar situation holds as well for the pseudoscalar FFs $P_1^s(w_{max}^*)$ and $P_1(w_{max}^*)$, due to the KC at maximum recoil. The origin of such differences between the cases of final pseudoscalar or vector mesons is unclear.

An interesting question is whether the spectator-quark dependence of the hadronic FFs shown in Figs.\,\ref{fig:FFMMBD}-\ref{fig:FFMMBDstar} is consistent with the experimental LHCb results on the ratios of branching fractions given in Eqs.\,(\ref{eq:R_exp})-(\ref{eq:Rstar_exp}). The latter ones do not depend on $\vert V_{cb} \vert$ and are affected by $SU(3)_F$ breaking effects within a $\simeq 10\%$ level. 
Using the DM bands of Figs.\,\ref{fig:FFMMBD}-\ref{fig:FFMMBDstar} we have evaluated the total decay rate modulo $\vert V_{cb} \vert^2$ for each of the four decay channels. Then, adopting the PDG values for the $B^0$ and $B_s$-meson lifetime, i.e.~$\tau_{B^0} = (1.519 \pm 0.004) \cdot 10^{-12} ~ \mbox{s}$ and $\tau_{B_s} = (1.516 \pm 0.006) \cdot 10^{-12} ~ \mbox{s}$\,\cite{ParticleDataGroup:2020ssz}, we obtain
\bea
    \label{eq:R_DM}
     \frac{\mathcal{B}(B_s \to D_s \mu \nu)}{\mathcal{B}(B \to D \mu \nu)} \Bigl|^{\rm DM} & = & 1.02 \pm 0.06 ~ , ~ \\[2mm]
    \label{eq:Rstar_DM}
    \frac{\mathcal{B}(B_s \to D_s^{*} \mu \nu)}{\mathcal{B}(B \to D^* \mu \nu)} \Bigl|^{\rm DM} & = & 1.19 \pm 0.11 ~ . ~ 
\eea
Within the present uncertainties of the order of $\approx 10 \%$ the above theoretical results agree with the corresponding experimental values\,(\ref{eq:R_exp})-(\ref{eq:Rstar_exp}) as well as with the updated results from Ref.\,\cite{LHCb:2021qbv}, which read
\bea
    \label{eq:R_exp_new}
     \frac{\mathcal{B}(B_s \to D_s \mu \nu)}{\mathcal{B}(B \to D \mu \nu)} & = & 1.04 \pm 0.09 ~ , ~ \\[2mm]
    \label{eq:Rstar_exp_new}
    \frac{\mathcal{B}(B_s \to D_s^{*} \mu \nu)}{\mathcal{B}(B \to D^* \mu \nu)} & = & 1.03 \pm 0.11 ~ . ~ 
\eea
A drastic improvement of the accuracy of both the experiments and the theory is mandatory in order to clarify the impact of $SU(3)_F$ breaking effects on both the branching fractions and the hadronic FFs.

Thus, further studies of the spectator-quark dependence of the hadronic FFs of the semileptonic $B \to D^{(*)}$ and  $B_s \to D_s^{(*)}$ transitions are called for. In particular, it would be very interesting to compare the dependence of the FFs on the spectator quark evaluated at fixed recoil on the same lattice gauge configurations.

\section{Conclusions}
\label{sec:conclusions}

In this work we have applied the Dispersive Matrix approach of Ref.\,\cite{Martinelli:2021onb,Martinelli:2021myh,DiCarlo:2021dzg} to describe the hadronic FFs of the exclusive semileptonic $B_s \to D_s^{(*)} \ell \nu_\ell$ decays in the whole kinematical range using the results of recent lattice QCD computations\,\cite{McLean:2019qcx, Harrison:2021tol} only at large values of the 4-momentum transfer. 
We have performed three analyses to extract $\vert V_{cb} \vert$ from the LHCb experimental data\,\cite{Aaij:2020hsi,LHCb:2020hpv,LHCb:2021qbv}, obtaining $\vert V_{cb} \vert \cdot 10^3 = 41.7 \pm 1.9$ from $B_s \to D_s \ell \nu_\ell$ and $\vert V_{cb} \vert \cdot 10^3 = 40.7 \pm 2.4$ from $B_s \to D_s^* \ell \nu_\ell$ decays.
After averaging with the values obtained from the $B \to D^{(*)}$ channels in Refs.\,\cite{Martinelli:2021onb,Martinelli:2021myh} our final result for $\vert V_{cb} \vert$ reads
\be
    \vert V_{cb} \vert^{\rm DM} \cdot 10^3 = 41.2 \pm 0.8 ~ \qquad \mbox{from $B_{(s)} \to D_{(s)}^{(*)} \ell \nu_\ell$ decays} ~ , ~ \nonumber
\ee
which is compatible with the most recent inclusive estimate, $\vert V_{cb} \vert_{\rm{incl}} \cdot 10^3 = 42.16 \pm 0.50$\,\cite{Bordone:2021oof}, at the $1 \sigma$ level.
We stress that in our bin-per-bin analysis the hadronic FFs (including their uncertainties) are determined exclusively by our fundamental theory of strong interactions, i.e.~QCD, while the experimental data are used only  to obtain the final exclusive determination of $\vert V_{cb} \vert$.

We have computed the theoretical values of the LFU ratios $R(D_s^{(*)})$ of the total decay rates in order to test LFU in these decays. Our results are
\be
    R(D_s) = 0.298~(5) ~ , ~ \qquad R(D_s^*) = 0.250~(6) ~ . ~ \nonumber
\ee

We have addressed also the issue of the $SU(3)_F$ symmetry breaking by comparing the hadronic FFs entering the semileptonic $B \to D^{(*)}$ and $B_s \to D_s^{(*)}$ channels. While the FFs relevant for transitions into pseudoscalar mesons have small $SU(3)_F$ breaking effects in the whole kinematical range, some of the FFs related to transitions into vector mesons exhibit larger $SU(3)_F$ breaking effects particularly at maximum recoil. This issue needs to be further investigated in the future by dedicated accurate LQCD simulations.

\begin{acknowledgments}
We warmly thanks Fabio Ferrari for fruitful discussions concerning the LHCb experiments of Refs.\,\cite{Aaij:2020hsi, LHCb:2020hpv} and Mirco Dorigo for drawing our attention to the updated LHCb results of Ref.\,\cite{LHCb:2021qbv}.
S.S.\,is supported by the Italian Ministry of Research (MIUR) under grant PRIN 20172LNEEZ.
\end{acknowledgments}

\bibliography{BsDs}
\bibliographystyle{JHEP}

\end{document}